\documentclass[aps,prd,reprint,11pt,onecolumn,amsmath,amssymb,nofootinbib,groupedaddress,superscriptaddress,floatfix,tightenlines,eqsecnum]{revtex4-2}

\usepackage{macros}

\begin{document}

\title{Big Bang For Your Helium Buck}

\author{Marilena Loverde}
\email{mloverde@uw.edu}
\affiliation{%
 Department of Physics, University of Washington, Seattle, WA, USA
}
\author{Murali M. Saravanan}
\email{msarav@uw.edu}
\affiliation{%
 Department of Physics, University of Washington, Seattle, WA, USA
}%
\author{Zachary J. Weiner}\email[]{zweiner@perimeterinstitute.ca}
\affiliation{Perimeter Institute for Theoretical Physics, Waterloo, Ontario N2L 2Y5, Canada}

\date{\today}

\begin{abstract}
We study the implications of a recent measurement of the primordial helium fraction from the Large
Binocular Telescope for cosmological inference from the cosmic microwave background.
We show that LBT establishes the robustness of cosmological parameters to theoretical assumptions
about big bang nucleosynthesis: its empirical calibration of the helium fraction enables constraints
on cosmology and inflation that are agnostic to BBN but as precise as those that instead enforce
standard BBN predictions.
Future CMB surveys require at most a marginal improvement in precision over LBT to maximize their
BBN-agnostic constraining power when the radiation density is free (and no more than a factor of two
improvement across all cases we consider).
We then apply the LBT measurement to a number of scenarios that feature new physics in BBN and the
CMB.
First, we constrain nonstandard radiation sectors (interacting and free-streaming) and search for evolution of the radiation
abundance between nucleosynthesis and recombination.
We then test models that alleviate the Hubble tension and the tension between CMB and
baryon acoustic oscillation data, including self-interacting light relics and varying fundamental
constants; LBT precludes most but not all of the models we consider via their effect on BBN.
Finally, we use the LBT measurement as an indirect but independent test of the neutron lifetime
anomaly, inferring values that are consistent with ``bottle'' experiments and $2.4 \sigma$ below
``beam'' experiments.
\end{abstract}

\maketitle
\makeatletter
% reduce spacing above each section entry
\let\oldl@section\l@section
\renewcommand{\l@section}[2]{\vspace{-0.3\baselineskip}\oldl@section{#1}{#2}}
% hide subsubsections
\def\l@subsubsection#1#2{}
% hide subsections
% \def\l@subsection#1#2{}
\makeatother
% reduce space above toc
\vspace{-\baselineskip}
\tableofcontents

%\tableofcontents
\section{Introduction}\label{sec:intro}
The cosmic microwave background (CMB) is presently the most pristine source of information on the
origin of structure and the matter contents of the Universe.
The great success of general relativity and the Standard Model of particle physics in predicting the
evolution of the plasma---from an early state of equilibrium through hydrogen recombination---has
enabled observations of CMB anisotropies to establish the dominant matter components of the
Universe, tightly limit new particle species, and confirm the Gaussian, red-tilted power spectrum of
primordial curvature perturbations predicted by the simplest inflationary
scenarios~\cite{WMAP:2012nax, Planck:2018vyg, AtacamaCosmologyTelescope:2025blo, SPT-3G:2025bzu}.
With the rich spatial information in its temperature and polarization anisotropies, the CMB is directly sensitive to all Standard Model processes relevant to the dynamics of the plasma, save the synthesis of light nuclei (and decoupling of neutrinos) at temperatures roughly five orders of magnitude above that of last scattering.
Big bang nucleosynthesis (BBN)~\cite{Iocco:2008va, Cyburt:2015mya} forms a second pillar of early universe cosmology that is both complementary to the CMB and implicit in many of its flagship results.
Knowledge of BBN is particularly important to the CMB's measurements of the baryon abundance,
searches for new light relic species, and constraints on the shape of the primordial power spectrum. 

The role of BBN in analyses of the CMB is in setting the density of free electrons and, in turn,
the rates that govern hydrogen recombination and the dynamics of the photon-baryon plasma.
A fraction of the electrons are first bound in helium atoms, which recombine before hydrogen.
Since the Universe is charge neutral and no other elements have appreciable primordial abundance,
the primordial helium fraction is the only result of primordial nucleosynthesis required to model
the CMB.
While the synthesis of helium during BBN is well understood within the Standard Model, new physics
(even if decoupled from the Standard Model) can modify the predicted helium
yield~\cite{Pospelov:2010hj, Cyburt:2015mya, Yeh:2022heq}.
Even within the Standard Model, the predicted light element abundances depend on the measured values
of the neutron lifetime~\cite{Iocco:2008va, Cyburt:2015mya, Chowdhury:2022ahn, Yeh:2023nve}, whose
empirical determination differs between experimental
methods~\cite{Yue:2013qrc,Fuwa:2024cdf,ParticleDataGroup:2026xqw}, and nuclear cross
sections~\cite{Pitrou:2018cgg, Pitrou:2020etk, Yeh:2020mgl, Pisanti:2020efz, Launders:2026ciu}, whose uncertainties are irrelevant to the helium fraction but are a significant source of 
discrepancy for other elements like deuterium~\cite{Fields:2019pfx,
Burns:2026wlw}.

A maximally conservative approach to constraining CMB-era cosmology could marginalize over any
possible modification of BBN by treating the helium yield, the only relevant input, as an
independent free parameter.
The cost of such BBN-agnostic analyses, compared to BBN-consistent ones that enforce standard
BBN predictions, is a substantial reduction in the precision with which a number of cosmological
parameters are inferred, including order-unity degradations for the baryon abundance, the scalar
spectral index, and the abundance of relativistic
species~\cite{Hou:2011ec,Planck:2018vyg,AtacamaCosmologyTelescope:2025nti,SPT-3G:2025bzu}.
Astronomical determinations of the primordial helium nucleon fraction, $\Yp$, provide an empirical route
to mitigate this loss in constraining power and test the robustness of the flagship results from the
CMB to assumptions about BBN.
Recently, the Large Binocular Telescope (LBT) project measured $\Yp$ with $0.5\%$
precision~\cite{Aver:2026dxv} using high-quality spectra of the ionized extragalactic medium of extremely metal-poor galaxies.
Their measurement improves in precision over the Particle Data Group's (PDG) current recommended
value~\cite{ParticleDataGroup:2026xqw} by a factor of 2.3, due in part to sample homogeneity and
improved emissivity and radiation transfer modeling.

In this work, we explore the implications of the LBT measurement for BBN-agnostic inference from the
CMB.
We show that, for current CMB data, empirically calibrating $\Yp$ with the LBT measurement yields
BBN-agnostic constraints on other cosmological parameters that match the precision obtained by
instead imposing standard BBN as a theoretical calibration.
The concordance of empirically and theoretically calibrated analyses establishes the robustness of
cosmological constraints to assumptions about BBN.
We then leverage BBN's independent value as a probe of new physics and study scenarios that uniquely
benefit from simultaneously modeling BBN and combining LBT and CMB data, including new physics in
the radiation sector and models of interest to current cosmological tensions.

Information on the helium fraction is especially important to constraining the possible existence of new light relic particles, which is a leading science driver for ongoing and future high-resolution CMB experiments, such as the South Pole Observatory~\cite{SPT-3G:2019sok}, Simons Observatory~\cite{SimonsObservatory:2018koc}, or a future CMB-S4-like survey~\cite{Trendafilova:2026xtu}.
The Standard Model radiation content after electron-positron annihilation entirely comprises photons and relativistic neutrinos, 
$\rho_r = \rho_\gamma + \rho_\nu$, whose energy densities are related by,
\begin{align}
    \rho_r
    = \rho_\gamma \left[ 1 + \frac{7}{8} \left( \frac{4}{11} \right)^{4/3} \Neff \right]
    .
\label{eqn:Neffdef}
\end{align}
The Standard Model (SM) predicts the effective number of neutrino species to be $\Neff = 3.044$,
including the effects of the noninstantaneous decoupling of the three neutrino
species~\cite{Mangano:2005cc, Akita:2020szl, Bennett:2020zkv, Froustey:2020mcq, Cielo:2023bqp},
and many models of physics beyond the Standard Model (BSM) predict additional contributions to or
other deviations from this baseline value (see Refs.~\cite{CMB-S4:2016ple,Dvorkin:2022jyg,Trendafilova:2026xtu} for reviews).
The uncertainty on $\Neff$ inferred from current CMB data (combined with baryon acoustic oscillation
[BAO] data) degrades from $0.11$ to $0.23$ due to BBN agnosticism. We show the precision on $\Neff$ can be fully restored to its
BBN-consistent value by including the LBT measurement. This qualitative result extends to CMB measurements of
the baryon density and inflationary parameters.

Beyond mitigating degeneracies in CMB predictions between $\Yp$ and cosmological parameters like
$\Neff$, light element abundances are uniquely valuable as a direct and CMB-independent probe of
new physics active during BBN~\cite{Cooke:2017cwo, Yeh:2026pil}.
In particular, combining LBT with measurements of the deuterium abundance (to determine the
baryon-to-photon ratio) yields $\Neff = 2.941 \pm 0.092$~\cite{Yeh:2026pil}, exceeding the precision
of the CMB constraints of $\Neff = 3.02 \pm 0.11$ (as obtained under theoretical or empirical
calibration of $\Yp$ as we show in \cref{sec:standard_cosmo}).
Moreover, modeling BBN enables joint constraints that, by combining theoretical and empirical
information on $\Yp$, exceed both in sensitivity but are not robust to assumptions about BBN; the current leading CMB--BBN constraint (when including BAO data) is 
$\Neff = 3.004\pm 0.071$~\cite{Goldstein:2026iuu}.

These joint constraints assume the same abundance of relativistic species at BBN and recombination,
but the helium yield and the CMB are sensitive to different eras, enabling searches for nontrivial
evolution in between~\cite{Yeh:2022heq}.
Moreover, unlike BBN, the CMB probes the dynamics of radiation
perturbations~\cite{Bashinsky:2003tk}, informing particle properties beyond just the homogeneous
abundance~\cite{Baumann:2015rya}.
We use the LBT measurement and CMB data to study these nonminimal scenarios in
\cref{sec:new_physics}.
In addition to constraining evolution in $\Neff$, we search for light relics that, in contrast to
the default assumption, have strong self-interactions.
We also perform a consistency test between the radiation density inferred from LBT and from the CMB
by marginalizing over the behavior of perturbations.

Measurements of the light element abundances can discriminate between explanations of
anomalies and tensions that also predict effects on BBN.
Though high-resolution CMB data have long excluded light relics as solutions~\cite{Wyman:2013lza,
Bernal:2016gxb} to the tension with local determinations of the Hubble constant~\cite{Riess:2016jrr,
Riess:2019cxk, Riess:2021jrx, Freedman:2024eph, H0DN:2025lyy}, dark radiation may still alleviate
the milder tension between CMB data and distances inferred from baryon acoustic
oscillations~\cite{DESI:2025zgx}, especially if strongly self-interacting~\cite{Allali:2024cji,
Saravanan:2025cyi}.
The LBT measurement precludes this possibility in scenarios where the contribution to the radiation
density is also present during nucleosynthesis.
Likewise, variation in the fundamental constants between recombination and the present could resolve
the Hubble~\cite{Hart:2019dxi, Sekiguchi:2020teg} and BAO--CMB tensions~\cite{Baryakhtar:2024rky,
Weiner:2026sfm}; minimal models of hyperlight scalar fields generally shift the constants at BBN
just as much as at recombination~\cite{Baryakhtar:2024rky, Baryakhtar:2025uxs}.
The LBT measurement is also precise enough to meaningfully weigh in on the discrepancy in
measurements of the neutron lifetime from different experimental methods~\cite{Yeh:2023nve}.

After briefly reviewing the theoretical interplay of nucleosynthesis and CMB anisotropies,
\cref{sec:cosmo} studies the impact of BBN agnosticism and the LBT measurement on standard
cosmological parameters and on inflationary parameters measured by current CMB data.
\Cref{sec:forecasts} forecasts the precision requirements of astrophysical measurements of $\Yp$
that would maximize BBN-agnostic constraints from future CMB observations.
We extend the standard analysis of light relics to a number of nonminimal scenarios in
\cref{sec:new_physics}.
Finally, in \cref{sec:tensions} we discuss the implications of the light element abundances for
tensions among cosmological datasets as well as the neutron lifetime anomaly.
We conclude in \cref{sec:conc}.
\Cref{sec:full_constraints} presents supplementary results, including extended parameter sets and
additional dataset combinations, and \cref{sec:methods} describes our methodology.

\section{Cosmology from the CMB agnostic to physics at nucleosynthesis}\label{sec:cosmo}

In this section, we compare cosmological constraints from the CMB that assume the standard theory of BBN with ones that are fully independent of the physics of BBN.
We demonstrate that the observational determination of the helium abundance from the LBT project~\cite{Aver:2026dxv} enables empirically calibrated measurements of cosmological parameters commensurate with theoretically calibrated ones, i.e., that model BBN instead.
We first review the dependence of the helium abundance prediction on cosmological parameters, the CMB's own sensitivity to light element abundances, and the role of theoretical assumptions and empirical measurements of the helium abundance in analyses of CMB data.
 
In the Standard Model, the primordial element abundances depend only on the baryon-to-photon number
ratio $\eta$~\cite{Grohs:2023voo,Cyburt:2015mya}.
For any viable cosmologies, virtually all the nucleons are synthesized into helium or remain free
protons (hydrogen nuclei).
Nucleosynthesis also depends on the total energy density deep in the radiation era (via the
expansion rate): with a higher radiation density, the weak interactions freeze out at earlier times, and fewer neutrons decay before the deuterium bottleneck clears,
leading to a higher neutron-to-proton ratio and thus a higher helium-4
abundance.
The helium abundance is therefore sensitive to not just the abundance of neutrinos (even after they
decoupled), but also any additional (decoupled) radiation beyond the Standard Model.
In standard BBN calculations, the primordial helium nucleon fraction
$\Yp \equiv 4\nHe/ n_\mathrm{B} \simeq 4\nHe/ \left(4\nHe + \nH\right)$ (with $n_\mathrm{B}$,
$\nHe$, and $\nH$ the number densities of baryons, helium nuclei, and hydrogen nuclei), is much more
sensitive to $\Neff$ than to $\eta$, being nearly independent of the latter.
An effective fitting function to $\Yp$ is
\begin{align}
    \Yp
    &= 1 - \left( 1 - 0.246956 \right) 
        \left( \frac{\eta}{6.104 \times 10^{-10}} \right)^{-0.0127}
        \left( \frac{\bar{\rho}_r / \bar{\rho}_\gamma}{1.691} \right)^{-0.1317}
    ,
    \label{eqn:abundance-scalings}
\end{align}
where $\bar{\rho}_r / \bar{\rho}_\gamma$ is the radiation-to-photon density ratio at BBN.
\Cref{eqn:abundance-scalings} implies that 
$\Yp \propto \eta^{0.039} \Neff^{0.17}$~\cite{Fields:2019pfx, Burns:2026wlw}.\footnote{
    Placing the parameter scaling on the hydrogen rather than helium yield provides a fit
    that remains accurate over a much broader parameter range, i.e., fitted exponents exhibit
    substantially less running.
    So far as we can tell, this improvement is a numerical accident; expressing the fit in
    terms of $\bar{\rho}_r$ instead of $\Neff$, on the other hand, more directly parametrizes the
    physics (namely, the expansion rate).
    We derive \cref{eqn:abundance-scalings} from results tabulated from 
    PRIMAT~\cite{Pitrou:2018cgg, Pitrou:2020etk} that are bundled in CAMB~\cite{Lewis:1999bs}.
    It is accurate to the $0.05\%$ level (an order of magnitude below LBT's precision) for 
    $1.5 \lesssim 100 \omega_b \lesssim 3.8$ (or $4.1 < 10^{10} \eta < 10$)
    and $1 \lesssim \Neff \lesssim 4$, more than sufficient for fitting to the LBT 
    data (and far more than sufficient for CMB data).
}

Of the light elements produced during BBN, helium and hydrogen are not just the only ones
synthesized in appreciable number but also the only ones that play an important role in the
generation of the CMB.
By parametrizing the relative amount of free electrons before and after helium recombination, $\Yp$
controls the ionization fraction (the ratio of free electrons to hydrogen nuclei) at the start of
hydrogen recombination~\cite{Hu:1995fqa,Bashinsky:2003tk}.
The ionization fraction in turn affects the recombination history and the Thomson scattering rate
when the primary features in CMB anisotropies are generated~\cite{Hu:1996mn}.
The resulting changes to the photon visibility function impact the generation of polarization and
the suppression of small-scale anisotropies from averaging over the visibility, while the Thomson
rate primarily affects diffusion damping, which additionally suppresses anisotropies on scales
smaller than the photon mean free path~\cite{Hou:2011ec}.

Though the default, ``BBN-consistent'' analyses of the CMB assume $\Yp$'s standard
dependence on $\eta$ and $\Neff$, ``BBN-agnostic'' analyses that instead treat $\Yp$ as an
independent free parameter still constrain its value, primarily through its impact on the damping
tail.
(Strictly speaking, a standard BBN-consistent analysis assumes that $\eta$ and $\Neff$ are the same, constant
value at BBN and last scattering.)
In either case, astronomical measurements of the primordial helium abundance provide independent
information that also assumes nothing about the physics at nucleosynthesis.
Such measurements therefore enable an ``empirically calibrated'' analysis in the BBN-agnostic case
to complement the ``theoretically calibrated'' one that is consistent with BBN physics but excludes the
empirical constraint.
In this work, we employ the LBT project's measurement, specified as a normal distribution
with mean $0.2458$ and standard deviation $0.0013$,
\begin{align} \label{eqn:LBT}
    \Yp \sim \mathcal{N}(0.2458, 0.0013).
\end{align}
Since baryons impact the CMB through their energy density, CMB calculations typically parametrize the
helium fraction by mass $\YHe$, related to the fraction by nucleon $\Yp$ by 
$\YHe = \Yp / \left( [1 -
\Yp] \cdot 4 \mH / \mHe + \Yp \right)$, where the mass ratio of four hydrogen nuclei to one helium
nucleus is $4 \mH / \mHe \approx 1.00717$~\cite{Pitrou:2018cgg}. 
The difference between the two is in fact the same size as the uncertainty of the LBT measurement; exchanging them can therefore bias inferred parameters by around $1 \sigma$.

In \cref{sec:standard_cosmo}, we study the impact of the LBT measurement on the standard cosmological model and models where the effective number of neutrino species is allowed to vary.
We show that empirically and theoretically calibrated analyses of CMB data now yield cosmological constraints of matching precision.
Given its consistency with the SM prediction, the LBT measurement therefore evidences the robustness of the standard cosmological parameters inferred from the CMB to removing all assumptions about the physics of BBN.
Employing the LBT measurement while enforcing consistency with BBN predictions for $\YHe$ yields even tighter constraints on $\Neff$~\cite{Yeh:2026pil,Goldstein:2026iuu}; however, we caution that the consistency of these measurements with the SM prediction $\Neff = 3.044$ is sensitive to the choice of datasets.
We note where dataset choice has an impact in \cref{sec:standard_cosmo}, leaving further elaboration to \cref{sec:new_physics,sec:full_constraints}.
Next, \cref{sec:inflation} examines the impact of $\YHe$ on the inference of inflationary
parameters, showing that the LBT measurement again enables BBN-agnostic constraints identical to
BBN-consistent ones.
Finally, in \cref{sec:forecasts}, we use idealized forecasts of high-resolution CMB observations to
assess the sensitivity of future cosmological constraints to assumptions about BBN, showing that the
precision provided by the LBT measurement is already nearly sufficient for maximally informative,
empirically calibrated results.

We employ the CMB temperature and polarization datasets from \Planck{} 2018
(PR3)~\cite{Planck:2019nip}, the Atacama Cosmology Telescope's DR6 release (ACT
DR6)~\cite{AtacamaCosmologyTelescope:2025blo}, and the South Pole Telescope's latest release (SPT-3G
D1)~\cite{SPT-3G:2025bzu}; throughout, we also utilize a joint lensing likelihood from the three
surveys~\cite{ACT:2025qjh}.
At times, we use baryonic acoustic oscillation (BAO) data from the Dark Energy Spectroscopic Instrument (DESI DR2)~\cite{DESI:2025zgx}.
\Cref{sec:methods} further details our methodology.
We assume a flat Universe and fix the summed mass of neutrinos to
$M_\nu \equiv \sum_i m_{\nu_i} = 60~\mathrm{meV}$, implemented as one massive neutrino species with all other light relics massless.
Finally, we denote the present-day energy density in each species $i$ by $\omega_i \equiv \rho_i 8\pi G/(3H_{100}^2)$, with $H_{100} \equiv 100~\mathrm{km}/\mathrm{s}/\mathrm{Mpc}$.

\subsection{Cosmological energy densities} \label{sec:standard_cosmo}

We begin with the standard \LCDM{} model and the extension where the effective number of neutrinos is a free parameter ($\LCDM{} + \Neff$).
In the Standard Model, $\YHe$ is a function of the baryon-to-photon ratio alone (setting aside other measured SM parameters like the neutron lifetime) and an extremely slowly varying one at that, such that $\YHe$ is essentially fixed in standard, BBN-consistent analyses.
Since both the helium fraction and the total baryon density determine the free electron fraction during recombination, excluding any constraint on $\YHe$ from BBN degrades the measurement of the baryon density.
\begin{figure}[t]
    \centering
    \includegraphics[width=\textwidth]{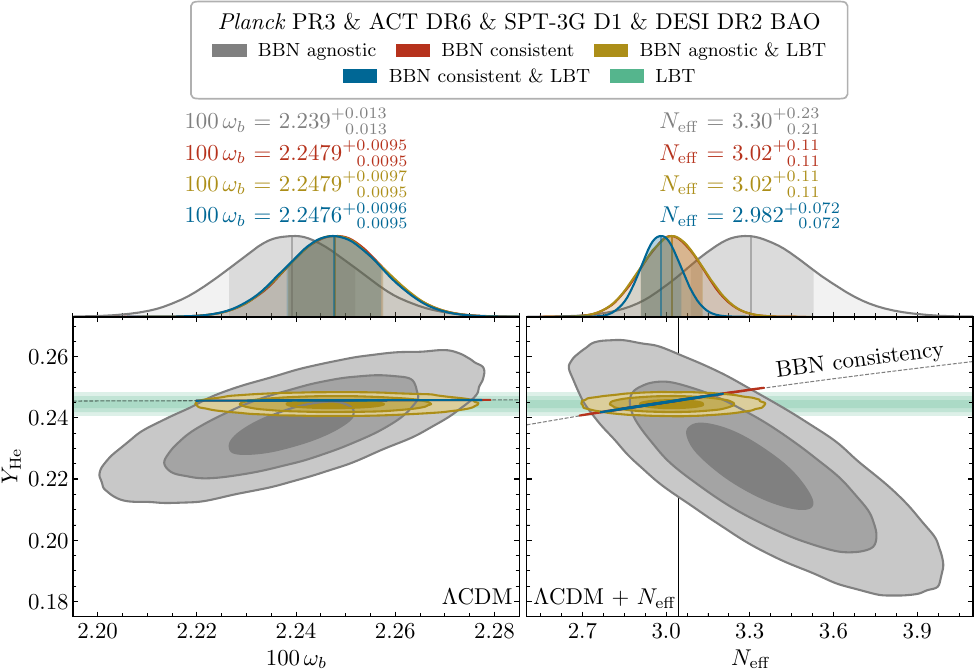}
    \caption{
        Joint posterior distributions of the primordial helium mass fraction $\YHe$ with the baryon density $\omega_b$ in \LCDM{} (left) and with $\Neff$ in $\LCDM{} + \Neff$ (right),
        deriving from CMB temperature, polarization, and lensing data from \Planck{}, SPT, and ACT as well as DESI DR2 BAO data.
        Dashed lines depict the parameter dependence of $\YHe$ in the standard model of BBN, as imposed for the red and blue posteriors (which respectively exclude and include the LBT measurement of $\YHe$).
        BBN-agnostic constraints (allowing $\YHe$ to vary freely; grey) are substantially less
        precise than the standard BBN-consistent results (red).
        While $\YHe$ is constrained by the theoretical predictions of standard BBN with drastically more precision than the empirical measurement from LBT (green), the latter is sufficient to recover the constraining power of BBN-consistent analyses.
        Namely, combining LBT with CMB and BAO data (gold) yields empirically calibrated
        measurements of $\omega_b$ and $\Neff$ that match the theoretically calibrated ones
        (evident in the complete overlap of the red and gold 1D posteriors in the top panels),
        making the results completely robust to theoretical assumptions about BBN.
        Taking both the LBT measurement and the standard model of BBN (blue) has no effect on
        $\omega_b$ (given $\YHe$'s negligible dependence on it)
        but does enable a more precise measurement of $\Neff$ (as obtained in Refs.~\cite{Yeh:2026pil,Goldstein:2026iuu}).
        Bottom panels display the 1, 2, and 3 $\sigma$ highest-density contours of the two-dimensional densities (i.e., the 39.3\%, 86.5\%, and 98.9\% mass levels); the shaded regions in the one-dimensional densities depict the $1 \sigma$ (i.e., equal-tailed $68\%$) regions, with medians marked by a vertical line. 
        The vertical line in the right panel marks the SM prediction of $\Neff = 3.044$. 
        Results for other cosmological parameters are presented in \cref{fig:lcdm_PAS,fig:Neff_PAS} in \cref{sec:full_constraints}.
    }
    \label{fig:diff_models_PAS}
\end{figure}
\Cref{fig:diff_models_PAS} shows that, while the CMB's measurement of $\omega_b$ is about $1.4$ times less precise when $\YHe$ is completely free, including the LBT measurement of $\YHe$ is sufficient to yield an identical posterior distribution to that from instead enforcing BBN consistency.
Evidently, $\YHe$ is \textit{predicted} in standard BBN far more precisely than needed to empirically mitigate its impact on the inference of other parameters from the CMB.
This finding extends to other \LCDM{} parameters which we discuss in more detail in \cref{sec:full_constraints}.

Allowing the effective number of neutrino species $\Neff$ to vary (as in the right panel of \cref{fig:diff_models_PAS}) yields an even more pronounced result, since $\Neff$ and $\YHe$ (independently varied) are both primarily constrained by their impact on the damping tail. 
However, the LBT $\YHe$ measurement once again sharpens the constraints on $\Neff$ to the same precision as enforcing BBN consistency.
\Cref{fig:Neff_PAS} in \cref{sec:full_constraints} shows that this result extends to cosmological parameters that are degenerate with $\Neff$ (and therefore also $\YHe$).
Since the radiation density, unlike other cosmological parameters, has significant effects on both the CMB and the prediction for $\YHe$, empirical measurements of $\YHe$ enable further improvements in the precision of $\Neff$ from the CMB when assuming standard BBN.
BBN specifically predicts an increase of $\YHe$ with $\Neff$ that exacerbates the change in the
diffusion scale, evident in \cref{fig:diff_models_PAS} as the BBN consistency line's rotation away
from the semimajor axis of the BBN-agnostic posterior.
Including the LBT measurement in a BBN-consistent analysis provides the most stringent determination of $\Neff$, yielding a $2.5\%$ measurement (consistent with Ref.~\cite{Goldstein:2026iuu}).

The measurement of $\Neff$ in \cref{fig:diff_models_PAS} is consistent with the SM prediction for neutrinos within $\lesssim 1 \sigma$.
This concordance is due in large part to the inclusion of DESI DR2 BAO data, which, though having only a marginal impact on the final precision, indirectly shifts the central values toward $3.044$
from the preference for $\Neff < 3.044$ from recent CMB datasets at roughly the $2 \sigma$ level.
DESI's effect reflects the extant tension between BAO and CMB data, which, if resolved through some independent means, would mitigate its contributed shift toward the SM value.
At the same time, without the CMB preference for smaller $\Neff$ (driven by ACT's measurements of the damping tail), DESI would instead drive a preference for $\Neff \gtrsim 3.044$.
We discuss the effect of these tensions further in \cref{sec:tensions-radiation,sec:full_constraints};
\cref{fig:violin_plot_N_x_PS_PAS_bao_lens} also provides a full summary of DESI's impact in a range of scenarios.

\subsection{Inflationary parameters} \label{sec:inflation}

In this section, we investigate the impact of the LBT measurement of $\YHe$ on the inference of inflationary parameters via the primordial curvature power spectrum,
\begin{align}
    \Delta_\mathcal{R}^2(k) 
    &= A_s \left( \frac{k}{k_\mathrm{p}} \right)^{n_s + \alpha_s \ln (k / k_\mathrm{p}) / 2 - 1}
    ,
\end{align}
with $A_s$, $n_s$, and $\alpha_s$ the amplitude, tilt, and running of the spectrum
and $k_\mathrm{p}$ a pivot scale conventionally fixed to $0.05~\Mpc^{-1}$.
Since the helium abundance affects the shape of the high-$\ell$ CMB power spectra via its impact on diffusion damping~\cite{Hou:2011ec}, measurements of the tilt and running of the primordial power spectrum degrade if $\YHe$ is not otherwise constrained.
Though inference of the tensor-to-scalar ratio $r$ is insensitive to assumptions about BBN, we jointly vary it in order to assess the implications for inflationary models~\cite{Planck:2018jri,Ellis:2023wic}.
We set the tensor tilt according to the consistency condition for single-field slow-roll inflation,
and $r$ is defined at the same pivot scale $k_\mathrm{p}$. 
Otherwise, we take the standard \LCDM{} cosmology and fix $\Neff$ to the SM prediction.

\begin{figure}[t]
    \centering
    \includegraphics[width = \textwidth]{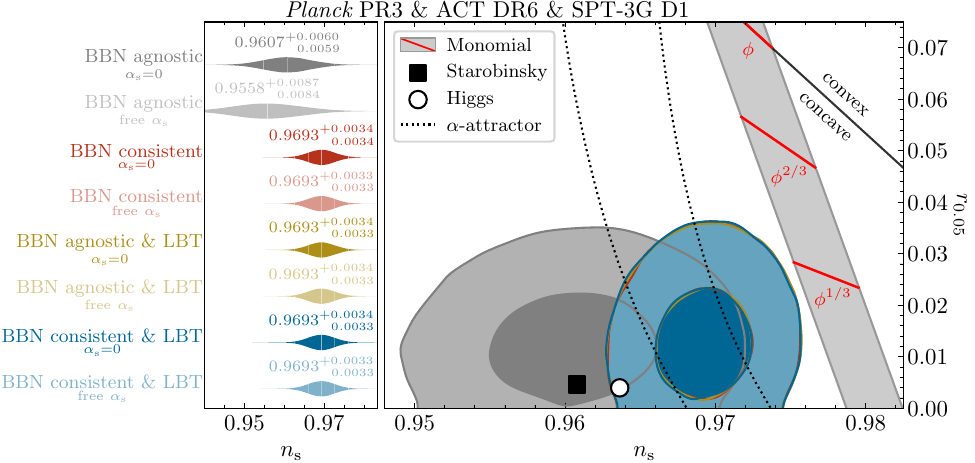}
    \caption{
        Dependence of the scalar spectral index $n_s$ and the tensor-to-scalar ratio $r$ measured by
        CMB data to assumptions about the primordial helium abundance $\YHe$, as in
        \cref{fig:diff_models_PAS}.
        Constraints use the \Planck{}, SPT, ACT, and BICEP/\textit{Keck} CMB temperature, polarization, and lensing data, 
        and the theoretical predictions of several representative inflationary models are displayed following Ref.~\cite{Balkenhol:2025wms}.
        While the constraint on $n_s$ shifts and broadens when fully agnostic to BBN (grey),
        its measurements are identical when $\YHe$ is either constrained by LBT (gold), set to the BBN prediction (red), or both (blue).
        The measurement of $r$ is robust across all cases.
        The left panel compares the constraints on $n_s$ for all combinations of BBN inputs
        with the running of the spectral index $\alpha_s$ fixed to zero (as taken in the right panel, solid colors) and freely varying (transparent colors); the robustness to BBN persists
        in this case.
    }
    \label{fig:r_n_s_PAS}
\end{figure}

\Cref{fig:r_n_s_PAS} compares constraints on the parameters of the primordial power spectra, with the running either fixed to zero or freely varied, under the same variations of BBN inputs considered in \cref{sec:standard_cosmo}.
We superimpose theoretical predictions for a representative set of reference models, following
Ref.~\cite{Balkenhol:2025wms} (see Refs.~\cite{Achucarro:2022qrl, Ellis:2023wic, Kallosh:2025ijd}
for recent reviews on inflation).
These models include monomial inflaton potentials, Starobinsky inflation, Higgs inflation, and polynomial $\alpha$-attractor models~\cite{Starobinsky:1980te,Mukhanov:1981xt,Starobinsky:1983zz,Bezrukov:2007ep,Bezrukov:2011gp,Kallosh:2022feu}.
We utilize the BICEP/\textit{Keck} B-mode CMB data~\cite{BICEP:2021xfz} in addition to the \Planck{}, SPT, and ACT CMB data used in \cref{sec:standard_cosmo}.
The current CMB--BAO tension shifts the parameters inferred from their combination in
\LCDM{}~\cite{Ferreira:2025lrd} ($n_s$ in particular); since these parameter shifts reflect a
compromise in fit that would be unnecessary if the tension were otherwise resolved, we take the
conservative approach of excluding the BAO data, as they do not directly constrain inflationary
parameters.

The measurement of $n_s$ depends strongly on the treatment of $\YHe$: even with $\alpha_s$ fixed
to zero, its precision degrades by nearly a factor of two when $\YHe$ is treated as a free
parameter.
Nevertheless, the evidence for a red-tilted (rather than scale-invariant) spectrum persists at the
$5$ to $6 \sigma$ level (compared to $\approx 9 \sigma$ in all other cases), due in part to a shift toward redder spectra.
As in \cref{sec:standard_cosmo}, including the LBT measurement of $\YHe$ is sufficient to restore the constraint on $n_s$ to exactly that from enforcing BBN consistency.
Because $n_s$ (in contrast to, e.g., $\Neff$ in \cref{fig:diff_models_PAS}) has no impact on the
predictions of BBN, results that do model BBN are unchanged by including the LBT measurement.
While varying $\alpha_s$ in addition to $\YHe$ degrades the uncertainty on $n_s$ even further
(and shifts it to even lower values), taking either the LBT measurement or the standard BBN prediction for $\YHe$ is sufficient to restore constraints identical to the case with 
$\alpha_s = 0$.
The three-way degeneracy responsible for this effect is depicted in \cref{fig:alpha_PAS} and discussed further in \cref{sec:full_constraints}; \cref{fig:alpha_PAS} also demonstrates the robustness of $\alpha_s$ to assumptions about BBN.
Finally, the limits on $r$ are dominated by the BICEP/\textit{Keck} B-mode data and are
robust to $\YHe$ as anticipated, whether or not $\alpha_s$ is varied (though the right panel of
\cref{fig:r_n_s_PAS} only displays results with $\alpha_s$ fixed to zero).

In summary, the empirical calibration of $\YHe$ from LBT is sufficiently precise to demonstrate the robustness of inflationary parameters inferred from the CMB to any assumptions about BBN.
This consistency precludes a BBN-based explanation of the $\gtrsim 2 \sigma$ exclusion of the Higgs and Starobinsky models depicted in \cref{fig:r_n_s_PAS}, which derives from ACT's preference for higher values of $n_s$ in \LCDM{}~\cite{Ferreira:2025lrd}.
That said, an alternative explanation of ACT's preference for reduced damping that is unconstrained by the $\YHe$ measurement could restore the CMB's compatibility with the lower values of $n_s$
measured by \Planck{} alone~\cite{Planck:2018vyg, Planck:2018jri}.

\subsection{Future BBN-agnostic constraints}\label{sec:forecasts}
\Cref{sec:standard_cosmo,sec:inflation} demonstrate that the LBT project's measurement of $\YHe$
establishes the robustness of standard cosmological parameters (both the abundances of all species
and inflationary parameters) to any assumption about the physics of nucleosynthesis.
We now assess whether the LBT measurement is sufficiently precise for this robustness between
empirically and theoretically calibrated analyses to persist with future CMB observations.
Cosmological parameters inferred with ongoing and proposed high-resolution experiments will depend
more and more dominantly on temperature and polarization data deep in the damping
tail~\cite{SimonsObservatory:2018koc}; a precise determination of $\YHe$, whether empirical or
theoretical, therefore becomes even more important to measurement precision.

\begin{figure}[t]
\includegraphics[width = \textwidth]{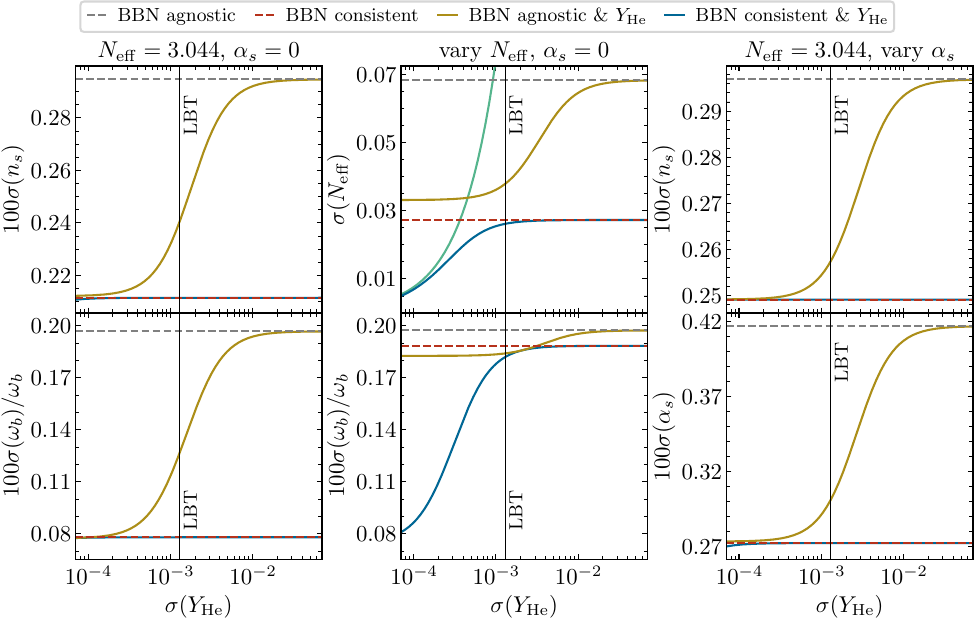}
\caption{
    Forecasted measurement precision for the baryon density $\omega_b$ and spectral tilt $n_s$ when
    fixing $\Neff$ to its SM prediction and the spectral running $\alpha_s = 0$
    (left), for $\Neff$ and $\omega_b$ when varying $\Neff$ (center),
    and for $n_s$ and $\alpha_s$ with $\Neff$ fixed (right).
    All forecasts take cosmic-variance limited CMB observations on $40\%$ of the sky up to
    a maximum multipole of $3000$ in temperature autocorrelation and $5000$ in polarization
    autocorrelation and the cross-correlation.
    Results are either agnostic to BBN (grey, dashed) or consistent with theoretical predictions for 
    BBN (red, dashed) or are agnostic (gold, solid) or consistent (blue, solid) while including
    external constraints on the helium abundance with uncertainty $\sigma(\YHe)$.
    The independent constraint on $\Neff$ from BBN alone for a given $\sigma(\YHe)$
    (taking an extremely conservative prior on $\omega_b$ with $2\%$ precision) appears in green.
    Vertical lines mark the precision of the current LBT measurement.
}
\label{fig:forecasts}
\end{figure}

To determine whether LBT's precision is sufficient for future surveys, we perform Fisher forecasts
for idealized CMB observations. 
Specifically, we forecast for cosmic-variance limited measurements of the unlensed temperature and polarization anisotropies for $\ell < 5000$ (and $\ell < 3000$ in TT) on $40\%$ of the sky~\cite{CMB-S4:2016ple}, neglecting lensing reconstruction as it has minimal interplay with $\Yp$ and the damping tail~\cite{Trendafilova:2026xtu,Bond:1997wr,Green:2016cjr}.
\Cref{fig:forecasts} compares likelihood configurations exactly analogous to those studied in \cref{sec:standard_cosmo,sec:inflation}.\footnote{
    For the most informative prior on $\YHe$ presented in \cref{fig:forecasts} (ten times more
    precise than the LBT measurement), the BBN-consistent forecasts are very marginally affected by
    propagating uncertainty in current measurements of the neutron lifetime $\tau_n$ (see
    \cref{sec:neutron-lifetime}).
    We ignore this uncertainty in \cref{fig:forecasts} for simplicity; only a modest improvement
    over the current leading measurement of $\tau_n$ would be required to mitigate its effect in this
    regime of extremely precise $\YHe$ measurements.
}
In \LCDM{} (with $\Neff$ fixed to the SM prediction), BBN agnosticism degrades the
forecasted measurement of $\omega_b$ by a factor of $2.5$ and of $n_s$ by $1.3$, which the LBT
result would mitigate in part but not in full.
A factor $\sim 2$ improvement in $\sigma(\YHe)$ would enable empirically calibrated constraints
to match theoretically calibrated ones.

Measurements of $\Neff$ are more robust than those of $\omega_b$ (when $\Neff$ is fixed): LBT's precision is
essentially sufficient to achieve a maximally precise BBN-agnostic constraint.
Even with an infinitely precise $\YHe$ measurement, however, a BBN-agnostic measurement comes just
short of the precision of a BBN-consistent one, since, as mentioned in \cref{sec:standard_cosmo},
the increase of $\YHe$ with $\Neff$ predicted by BBN amplifies the change in the damping rate.
In fact, measurements more precise than LBT's enable a BBN-consistent analysis to achieve greater
precision in both $\Neff$ and $\omega_b$, as the $\YHe$ measurement starts to dominate the
information on $\Neff$; however, only $\sigma(\YHe) < 10^{-4}$ is sufficient to fully break the
$\Neff$-$\omega_b$ degeneracy.
Otherwise, the $\Neff$-marginalized constraints on $\omega_b$ degrade only slightly from BBN
agnosticism.
Finally, the forecasted precision for inflationary parameters (the spectral tilt and running) is
nearly saturated in BBN-agnostic scenarios with a prior on $\YHe$ as precise as LBT's current
measurement.

\section{Implications for neutrinos and dark radiation}\label{sec:new_physics}

While the LBT measurement enables cosmological analyses to be agnostic to the physics at
nucleosynthesis with minimal loss in constraining power, the physics of BBN is well understood and
can be jointly leveraged to constrain BSM physics.
In particular, new light relics generically modify both CMB and BBN predictions, enabling stronger
joint constraints (as evident in \cref{fig:diff_models_PAS}); in this section, we consider a number
of extensions to the standard scenario of free-streaming light relics.
Given the tension between CMB datasets and DESI BAO data~\cite{DESI:2025zgx} (see
\cref{sec:tensions}), we exclude BAO data in this section but present analogous results that include
them in \cref{sec:full_constraints}.
Throughout this section we compare two dataset combinations: \Planck{} and SPT, and \Planck{}, SPT, and ACT with \Planck{} restricted to low multipoles
(\cref{sec:methods}).
Differences between the two therefore reflect both the addition of ACT data and the removal of high-$\ell$ \Planck{} data.
However, we include lensing data from all three experiments throughout this
paper as they slightly sharpen constraints without meaningfully shifting them~\cite{ACT:2025qjh}. 

In \cref{sec:dNfs_dNfld}, we explicitly constrain the abundance of new free-streaming or fluidlike
light relics on top of SM neutrinos, highlighting the nontrivial impact of recent CMB datasets as well as the LBT measurement.
Unlike light element abundances, the CMB is uniquely sensitive to properties of light relics beyond
their density; we therefore also derive $\Neff$ measurements that allow an arbitrary mixture of
noninteracting and self-interacting species for a consistency test with LBT's constraint on the radiation density alone.
As a more nontrivial application of jointly modeling BBN and applying the LBT measurement, in
\cref{sec:bbn_vs_cmb} we constrain the possible evolution of the radiation abundance ($\Neff$) between
nucleosynthesis and decoupling.

\subsection{Free-streaming and fluidlike light relics}\label{sec:dNfs_dNfld}

Dark radiation need not have the same macroscopic cosmological properties as Standard
Model neutrinos, which are entirely noninteracting after they decouple from the plasma.
A dark sector with efficient self-interactions remains tightly coupled and behaves like a perfect fluid rather than a gas of collisionless particles~\cite{Jeong:2013eza,Buen-Abad:2015ova,Chacko:2015noa}.
In some models, the efficiency of self-interactions can evolve on cosmological timescales, realizing transitions between these two limits~\cite{Cyr-Racine:2013jua, Choi:2018gho, Brinckmann:2022ajr}; here we restrict to the simpler possibility of relics that are always either free-streaming or fluidlike.
The perturbations of free-streaming radiation (like SM neutrinos) have distinct effects on the CMB anisotropies, modulating the overall height and phase of the acoustic oscillations~\cite{Bashinsky:2003tk,Baumann:2015rya}.
The degree of these distinguishing effects is determined by the fraction of the total radiation density that free streams, which is increased or decreased by new free-streaming or fluidlike relics, respectively.

\begin{figure}[t]
    \centering
    \includegraphics[width = \textwidth]{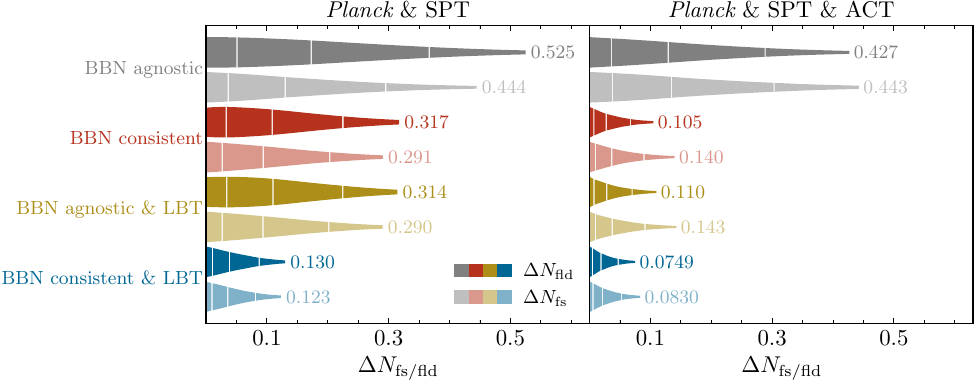}
    \caption{
        Posterior distributions for light degrees of freedom beyond the Standard Model neutrinos that are either free-streaming ($\Delta \Nfs$, transparent) or fluidlike ($\Delta \Nfld$, opaque).
        Results use the \Planck{} PR3 and SPT-3G D1 CMB temperature and polarization datasets,
        excluding (left) and including (right) ACT DR6 data.
        All results use CMB lensing data from all three surveys.
        Combinations of assumptions about BBN and of including the LBT measurement or not are indicated by color.
        As in \cref{sec:standard_cosmo}, while the LBT measurement yields constraints to the same precision as enforcing BBN consistency, the most stringent upper bounds on dark radiation require both empirically and theoretically calibrating $\YHe$.
        ACT, on the other hand, severely restricts the presence of light relics due to its nearly $2 \sigma$
        preference for $\Neff$ below the SM prediction~\cite{AtacamaCosmologyTelescope:2025nti}, independent of LBT (see
        \cref{fig:violin_plot_N_x_PS_PAS_lens}).
        ACT also more strongly penalizes fluidlike relics, due to the stringent upper limit it places on the angular size of the sound horizon $\theta_s$ (see \cref{fig:dNfs_dNfld_A_s_theta_s_lite} in \cref{sec:full_constraints}).
        Posteriors are truncated at the 95th percentile (whose value is also labeled); vertical white lines indicate the median and $\pm 1 \sigma$ quantiles.
    }
    \label{fig:dNfs_dNfld}
\end{figure}

\Cref{fig:dNfs_dNfld} presents constraints on light relics, fixing a minimum value of $\Neff = 3.044$ to account for SM neutrinos and allowing for additional free-streaming ($\Delta \Nfs$) or fluidlike ($\Delta \Nfld$) radiation.
As in \cref{sec:standard_cosmo}, an empirical calibration of $\YHe$ from the LBT experiment enables constraints as stringent as the theoretical calibration, making these constraints robust to any assumptions about BBN.
The upper limits on $\Delta \Nfs$ in \cref{fig:dNfs_dNfld} that are both consistent with BBN and use the LBT measurement (and that use all CMB data) are the tightest and are consistent with the results reported in Ref.~\cite{Goldstein:2026iuu}.

Comparing the limits on free-streaming and fluidlike radiation in \cref{fig:dNfs_dNfld} suggests
that recent CMB temperature and polarization data have a nontrivial impact, one beyond the precision gains expected from improved measurements of the damping tail.
In particular, while \Planck{} and SPT data allow for a greater abundance of fluidlike than of
free-streaming radiation, as previously found for \Planck{} data~\cite{Brust:2013ova,Baumann:2015rya,Blinov:2020hmc,Saravanan:2025cyi,Buen-Abad:2017gxg}, adding
ACT data tightens bounds on $\Delta \Nfld$ more so than $\Delta \Nfs$.
Most of the reduction derives from the $\sim 2 \sigma$ tension between ACT's damping tail and
\LCDM{} predictions rather than the improvement in precision.
However, since equal amounts of fluidlike and free-streaming radiation have the same effect on diffusion
damping (i.e., through the background expansion rate), ACT must also drive
shifted preferences for the fraction of all radiation that freely streams.

Weaker limits on fluidlike radiation are nominally expected on the grounds that it has a smaller differential impact on the free-streaming fraction
$\ffs \equiv \bar{\rho}_\mathrm{fs} / \bar{\rho}_r$ than free-streaming radiation~\cite[Eqn.
2.16]{Saravanan:2025cyi}, incurring a smaller associated shift in CMB height and phase.
Moreover, oscillatory effects incurred by the phase shift from fluidlike relics and the correlated
change in the baryon fraction (associated with the increase in radiation density) partly cancel,
leaving the overall effect more easily absorbed by a shift in the amplitude of the primordial power
spectrum~\cite{Saravanan:2025cyi,Ge:2022qws}.
As elaborated on in \cref{sec:full_constraints}, \cref{fig:dNfs_dNfld_A_s_theta_s_lite}
demonstrates that ACT disfavors a reduction in phase shift (i.e., from the reduction in $\ffs$ by
adding fluidlike radiation) strongly enough to oppose \Planck{}'s modest preference.
Regardless of the intrinsic preferences of the various CMB datasets, the BBN-consistent constraints
in \cref{fig:dNfs_dNfld} that also include the LBT measurement are nearly identical for fluidlike
and free-streaming relics.
The background-level information on the radiation density, from both the CMB damping tail and the
$\YHe$ measurement, therefore dominates over the perturbation-level signatures in CMB anisotropies.
Nevertheless, the phase and amplitude shifts are important distinguishing signatures of light
relics~\cite{Bashinsky:2003tk, Baumann:2015rya}, and the nontrivial preferences of recent datasets merit closer scrutiny, which we defer to future work.

Since the signatures of light relics in CMB anisotropies differ if they
interact efficiently, marginalizing over the behavior of their perturbations provides a more consistent comparison with constraints on their density from light element abundances.
We therefore extend the above analysis to a mixture of species that free stream (like SM
neutrinos) or are fluidlike~\cite{Saravanan:2025cyi}.
In marginalizing over the composition of radiation, we allow the effective degrees of freedom in
free-streaming radiation to be larger or smaller than $3.044$, meaning for part of the parameter
space the neutrinos are implicitly nonstandard.
While some of the parameter space might therefore not map straightforwardly into microphysical
models, and while dynamics may differ for dark radiation that interacts with matter components, this prescription provides a simple, phenomenological method to marginalize over a range
of qualitatively different behaviors of neutrinos and dark radiation~\cite{Saravanan:2025cyi}.

\begin{figure}[t]
\includegraphics[width = \textwidth]{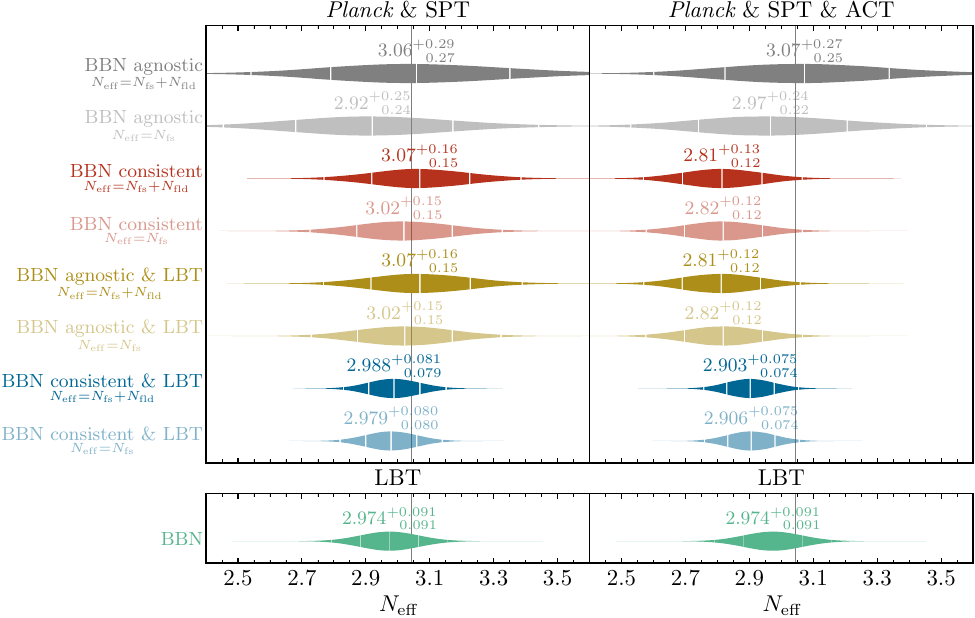}
\caption{
    Posterior distributions on the total radiation density parametrized by $\Neff$ [\cref{eqn:Neffdef}], using primordial element abundances and/or CMB datasets with different modeling choices for $\YHe$.
    Results use \Planck{} PR3 and SPT-3G D1 temperature and polarization data 
    and either exclude (left) or include (right) ACT DR6 data. 
    All results include lensing data from all three experiments.
    For the CMB posteriors, opaque results marginalize over the effect of radiation perturbations by allowing a mixture of free-streaming and fluidlike radiation (ignoring SM predictions for the neutrino sector).
    Transparent posteriors take all radiation to be free-streaming (labeled as $\Nfs$).
    Results either freely vary $\YHe$ (grey and gold) or set it according to BBN predictions (red and blue) and exclude (grey and red) or include (gold and blue) the LBT measurement.
    Posteriors are labeled by their median and 1$\sigma$ quantiles, with
    corresponding white lines indicating the median, 1$\sigma$, and 2$\sigma$ quantiles. 
}
\label{fig:violin_plot_N_x_PS_PAS_lens}
\end{figure}

\Cref{fig:violin_plot_N_x_PS_PAS_lens} presents constraints on $\Neff$ derived with the same
combinations of theoretical and empirical information on $\YHe$ (as in \cref{sec:standard_cosmo}),
comparing results for fully free-streaming radiation with cases that marginalize over the
free-streaming fraction.\footnote{
    When reporting BBN-only measurements of $\Neff$, as in \cref{fig:violin_plot_N_x_PS_PAS_lens}, we include the PDG deuterium abundance measurement~\cite{ParticleDataGroup:2026xqw} in order to break $\Yp$'s extremely weak degeneracy between
    the baryon-to-photon ratio $\eta$ and $\Neff$; that said, measurements of $\Neff$ are quite insensitive to this choice.
}
Marginalization only has a noticeable effect when the helium abundance is unconstrained by theory or
light element abundances; the CMB prefers a slightly higher density in radiation when not fully free
streaming. 
However, empirical and/or theoretical calibration of $\YHe$ drives measurements of the total radiation of equal precision, regardless of marginalization over the free-streaming fraction.
Just as when constraining additional radiation, the background information from the CMB's damping tail combined with $\YHe$ information dominates.
In particular, including ACT data makes the central values identical between marginalized and unmarginalized cases, with its preference for less damping driving a preference for radiation nearly 2$\sigma$ below the SM prediction.

\subsection{Evolving radiation abundance between nucleosynthesis and recombination}\label{sec:bbn_vs_cmb}

Light element abundances and the CMB are sensitive to the energy content of the Universe in
different eras, enabling their combination to search for nontrivial dynamics of radiation in the
interim period~\cite{Yeh:2022heq}.
In this section, we derive phenomenological constraints on the evolution of radiation using the
CMB to measure the effective number of relativistic degrees of freedom at recombination ($\NeffCMB$)
and LBT or CMB inferences of $\YHe$ for that at BBN ($\NeffBBN$).
Although both observables depend on dynamics over a few decades of expansion, for simplicity we take
the radiation abundance (i.e., $a^4 \bar{\rho}_r$) to be separately time independent over each
epoch.
This simple parametrization is sufficient to search for a wide array of BSM physics: a relativistic
dark species becoming nonrelativistic between BBN and recombination or vice
versa~\cite{Steigman:2013yua,Berlin:2017ftj,Yeh:2022heq}, energy injection into the
photon bath~\cite{Hu:1992dc,Poulin:2016anj,Sobotka:2022vrr}, or nonstandard neutrino
physics~\cite{Hasegawa:2019jsa,Berlin:2019pbq,Escudero:2026mgw}.

CMB anisotropies depend not only on $\NeffCMB$ but also $\NeffBBN$ through its impact on $\YHe$.
Since the helium abundance is the only imprint of nucleosynthesis on the CMB and is essentially
uniquely determined by $\NeffBBN$ in the class of scenarios we consider, the BBN-agnostic analyses
of \cref{sec:cosmo,sec:dNfs_dNfld} may alternatively be reinterpreted as BBN-consistent CMB constraints
on $\NeffCMB \neq \NeffBBN$.
That said, when marginalized over $\Neff$ (i.e., $\NeffCMB$), the CMB is consistent with values of
$\YHe$ even smaller than can be achieved in standard BBN by reducing $\NeffBBN$;\footnote{
    Here ``standard BBN'' denotes the extrapolation in the standard tabulation of 
    $\Yp(\omega_b, \Neff)$ to $\Neff < 3.044$ by simply decrementing the background energy density
    without attending to any secondary (nongravitational) effects.
} we therefore perform dedicated parameter inference that samples over $\NeffCMB$ and $\NeffBBN$,
with the added benefit of enforcing a flat prior over $\NeffBBN$ (versus the Jacobian-induced prior
from sampling uniformly in $\YHe$).
\begin{figure}[t]
\centering
\includegraphics[width = \textwidth]{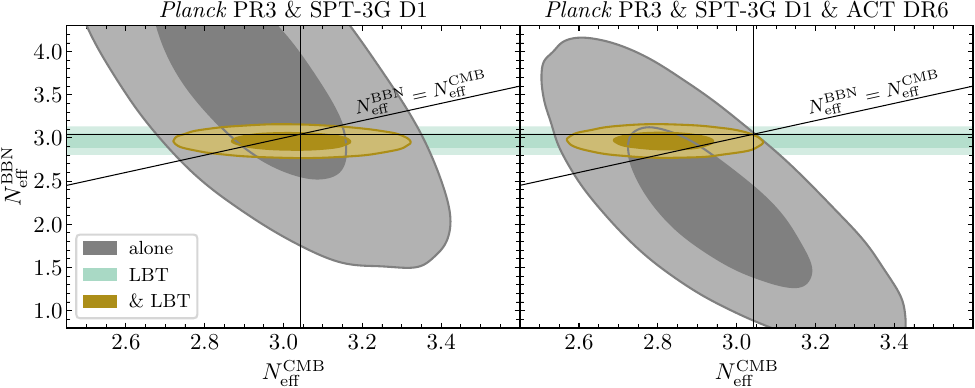}
\caption{
    Constraints on the evolution of the radiation abundance between BBN and CMB eras.
    Results use the \Planck{} PR3 and SPT-3G D1 CMB temperature and polarization datasets,
    excluding (left) and including (right) ACT DR6 data.
    All results use CMB lensing data from all three surveys.
    Black lines mark where $\NeffBBN = 3.044$, $\NeffCMB = 3.044$, and $\NeffBBN = \NeffCMB$.
    While \Planck{} and SPT data maintain consistency with $\NeffCMB = \NeffBBN = 3.044$, 
    the inclusion of ACT data drives preference for $\NeffBBN < 3.044$ and/or $\NeffCMB < 3.044$ when the CMB data are analyzed alone. 
    The LBT measurement requires $\NeffBBN$ to be consistent with the SM to $1 \sigma$, precluding
    any CMB preference for nontrivial evolution that modifies BBN. 
    See text for further discussion of implications of the constraints in the various quadrants for BSM physics.
}
\label{fig:deltaN}
\end{figure}
The CMB data from all three experiments contributes almost no information to the joint CMB--LBT measurement of 
$\NeffBBN = 2.964^{+0.096}_{-0.096}$, which is nearly identical to that from LBT 
($\NeffBBN = 2.974 \pm 0.091$).
\Cref{fig:deltaN} demonstrates that the constraints on $\NeffCMB$ depend on the combination of CMB datasets.
Results that include ACT data motivate BSM models featuring 
$\NeffCMB < \NeffBBN \simeq 3.044$~\cite{Escudero:2026mgw}, since LBT is consistent with the
Standard Model to within $1 \sigma$ and, as discussed in \cref{sec:dNfs_dNfld},
empirically calibrated CMB constraints on $\NeffCMB$ that include ACT data are nearly $2 \sigma$
lower than the SM prediction; constraints on $\NeffCMB$ with only \Planck{} and SPT data are
concordant with the SM prediction.

From a model-building perspective, values of $\NeffBBN>3.044$ and/or $\NeffCMB>3.044$ are the simplest to accommodate without modifying SM physics, i.e., with a secluded dark sector.
The simplest (and standard) case, $\NeffBBN = \NeffCMB > 3.044$, could be simply realized by some new light relic whose abundance was set at temperatures $T \gg \MeV$.
Less minimal scenarios could achieve unequal values (each still greater than the SM
prediction), $\NeffBBN > \NeffCMB > 3.044$, for instance if a fraction of the dark radiation becomes
nonrelativistic in between the two eras and is a portion of the dark
matter~\cite{Yeh:2022heq}.\footnote{
    Dark matter that becomes nonrelativistic relatively late generates various cosmological and 
    astrophysical signatures~\cite{Lin:2023fao}, making it challenging for the dark matter to
    make a sizeable contribution to $\NeffBBN$.
    In the case we consider, however, there is no need for the relativistic species to make up all of 
    dark matter; such a warm subcomponent would then have to become nonrelativistic rather close to 
    equality, making the transition in principle detectable in the CMB.
}
If the species equilibrates with the neutrinos after BBN and then becomes nonrelativistic, the reverse inequality
$\NeffCMB > \NeffBBN > 3.044$ could be realized via the resulting entropy transfer~\cite{Berlin:2017ftj,Berlin:2018ztp,Berlin:2019pbq,Krnjaic:2020znf},
though a separate contribution to $\Neff$ may be required to have extra radiation at BBN
($\NeffBBN > 3.044$).

The posteriors in the $\NeffBBN$--$\NeffCMB$ plane, however, have very little support in the area
of parameter space described above when ACT data are included, requiring $\NeffCMB<3.044$.
An effective number of neutrino species below the SM prediction necessitates a modification to Standard
Model physics via a reduction in the neutrino--photon temperature (or density) ratio.
This class of scenarios, unlike secluded dark sectors, thus necessarily features nongravitational
effects that are not negligible a priori and whose model dependence is outside of
the scope of this work (for a recent discussion, see Ref.~\cite{Escudero:2026mgw}).
In general, attempts to reduce the neutrino density (by having neutrinos decouple earlier) face stringent limits 
from terrestrial experiments as they require nonstandard neutrino interactions~\cite{Farzan:2017xzy,Mangano:2006ar,deSalas:2016ztq,Froustey:2026klr}.

Entropy injection into the photons, on the other hand, generates both strongly constrained CMB
spectral distortions~\cite{Hu:1992dc, Poulin:2016anj, Sobotka:2022vrr} and a variation in the
baryon-to-photon ratio that is limited by the combination of CMB and deuterium abundance
observations~\cite{Sobotka:2022vrr, Joseph:2026gws}.
(Since $\YHe$ is hardly sensitive to the baryon-to-photon ratio and we do not include deuterium
measurements, our analysis is insensitive to the correlated evolution in $\eta$.)
Though some of the parameter space may be challenging to realize microphysically (and may require
modified physics beyond shifts in $\Neff$ at each epoch), this analysis provides a straightforward
extension of the BBN-agnostic constraints studied in \cref{sec:standard_cosmo,sec:dNfs_dNfld}.
Ultimately, ACT data combined with the LBT measurement's consistency with the SM point toward new
physics that mostly affects the CMB (i.e., achieving $\NeffCMB < 3.044$, should this preference persist)~\cite{Escudero:2026mgw},
while \Planck{} and SPT data are consistent with the SM in both dimensions; secondary effects could
have interesting consequences for the fit to CMB data.

Finally, since the empirically calibrated constraints in \cref{fig:violin_plot_N_x_PS_PAS_lens} are agnostic to conditions at BBN, the comparison
of those results to the constraints on $\Neff$ from primordial abundances
generalizes the search for evolution in $\Neff$ from \cref{fig:deltaN} (up to priors) to marginalize
over the fraction of radiation that free streams as well.
Although we have excluded the DESI data to simplify the analysis throughout this section, including DESI DR2 BAO data significantly opposes the ACT-driven
preference for $\NeffCMB < 3.044$.
We discuss this further in \cref{sec:tensions-radiation} and present constraints on additional radiation as well in the $\NeffBBN$--$\NeffCMB$ plane in \cref{sec:full_constraints}.

\section{Implications for tensions and anomalies}\label{sec:tensions}

We next assess the implications of LBT's helium fraction measurement for solutions to cosmological
tensions and for the neutron lifetime anomaly.
\Cref{sec:cosmo-tensions} studies models that seek to explain tensions between \LCDM{} fits to
the CMB and low-redshift distances from baryon acoustic oscillations~\cite{DESI:2025zgx} and from
the calibrated distance ladder~\cite{Riess:2016jrr, Riess:2019cxk, Riess:2021jrx, Freedman:2024eph,
H0DN:2025lyy}.
\Cref{sec:neutron-lifetime} then shows that LBT can arbitrate the $\approx 6 \sigma$
disagreement in measurements of the neutron lifetime.

\subsection{Solutions to the Hubble and BAO--CMB tensions}\label{sec:cosmo-tensions}

In a variety of scenarios, new physics relevant at recombination also predicts changes at
nucleosynthesis, which light element abundance measurements can independently test.
For instance, in the simplest scenarios, extra density in the form of light relics (e.g., dark
radiation; \cref{sec:tensions-radiation}) is generically present at both nucleosynthesis and
recombination despite the five decades of expansion separating them: a standard, light thermal relic
must decouple far earlier than nucleosynthesis so that its contribution to $\Neff$ is not
far larger than is viable observationally (i.e., not order unity)~\cite{Dvorkin:2022jyg}.
Shifts in the early-time (prerecombination) values of the electron mass or fine-structure constant
(\cref{sec:varying-constants}) likewise generally apply at nucleosynthesis as well, as the hyperlight
scalar fields that realize the postrecombination transition to the present-day values remain frozen
at their initial misalignment at higher temperature~\cite{Baryakhtar:2024rky, Baryakhtar:2025uxs}.
Light element abundances therefore independently test such scenarios when consistently modeling
their effect on BBN.
We consider only the LBT measurement of $\YHe$ for its substantially greater observational precision,
theoretical robustness, and insensitivity to the baryon-to-photon ratio compared to the deuterium
abundance; we comment on recent studies of the impact of deuterium measurements in
\cref{sec:deuterium}.

\subsubsection{Light relics}\label{sec:tensions-radiation}

Light relics have long been considered as possible explanations of the Hubble
tension~\cite{Wyman:2013lza, Bernal:2016gxb}; while the possibility is well excluded by current CMB
data, they remain potential solutions to the BAO--CMB tension~\cite{Allali:2024cji,
Saravanan:2025cyi, Weiner:2026sfm}.
Increasing $\Neff$ above the SM prediction raises the expansion rate at recombination, which reduces
the sound horizon and increases the Hubble constant required to fit the angular locations of the
acoustic peaks.
Moreover, the matter density scales sublinearly with the radiation density in the CMB, due to
competing effects between preserving matter-radiation equality and the fraction of
pressure-supported matter~\cite{Saravanan:2025cyi}.
By decreasing the sound horizon to a greater degree than it increases the expansion rate in the
matter-dominated era, a new light relic alleviates the matter-era distance
excess~\cite{Weiner:2026sfm}, which incurs a correlated reduction in the CMB's predicted matter
fraction toward the values preferred by DESI.

\begin{figure}[t]
\includegraphics[width=\textwidth]{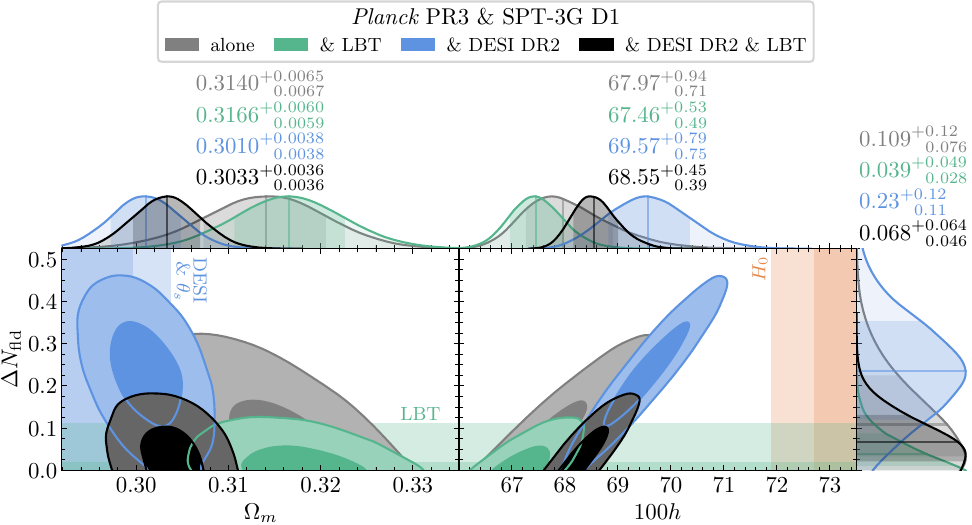}
\caption{
    Impact of LBT's helium abundance measurement on fluidlike light relics as solutions
    to cosmological tensions.
    Extra radiation increases the Hubble constant $h$ inferred by the CMB (grey) by increasing the
    expansion rate at recombination, thereby decreasing the sound horizon, but not nearly enough to
    resolve the tension with local determinations (orange)~\cite{H0DN:2025lyy}.
    The variations in the baryon and CDM density that preserve the CMB fit in response to the
    light relic's impact on the dynamics of perturbations also correlate to a reduction in the
    matter fraction $\Omega_m$, which in turn resolves the tension with DESI BAO data (represented
    in the vertical blue band by $\Omega_m$ measured by DESI and the CMB's analogous geometric
    information in $\theta_s$).
    The combination of CMB data and DESI (blue contours) yields a $\approx 2 \sigma$ preference
    for a fluidlike light relic.
    The LBT measurement of $\Yp$ independently constrains new light relics that are also present
    at BBN (horizontal green bands), which, combined with the CMB data (green contours) precludes
    any substantial alleviation of either tension.
    Combining both DESI and LBT with CMB data (black) modestly shifts $h$ and $\Omega_m$, but not
    meaningfully more so than the analogous \LCDM{} fit that simply compromises the fits to
    each dataset.
    Here we exclude the ACT DR6 CMB dataset, as its preference for a lower value of the radiation density
    than the SM prediction strongly disfavors new light relics of any kind; see
    \cref{fig:relic-tensions-violin} for a comparison.
    However, lensing data from all three surveys are used for these constraints.
}
\label{fig:relic-tensions}
\end{figure}

\begin{figure}[t]
\includegraphics[width=\textwidth]{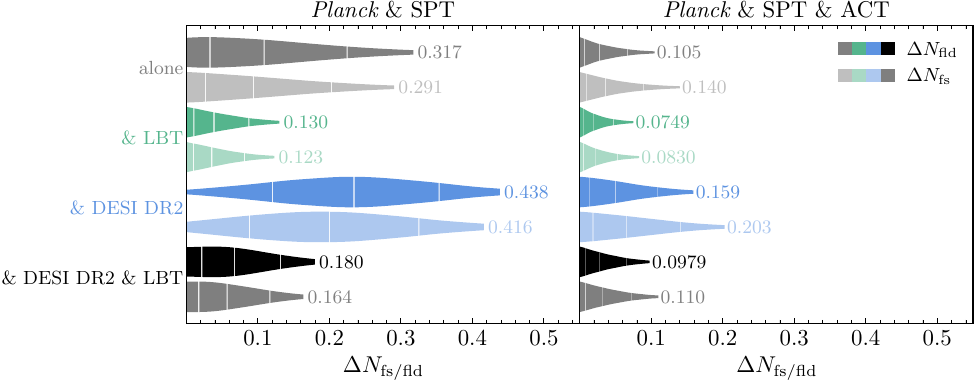}
\caption{
    Impact of LBT and DESI data on constraints on additional free-streaming ($\Delta \Nfs$, transparent) and fluidlike radiation ($\Delta \Nfld$, opaque), as derived with CMB data from
    \Planck{} PR3 and SPT-3G without and with ACT DR6 (left and right panels, respectively).
    All results use lensing data from all three experiments.
    Results that use CMB data alone (grey) and combined with the LBT measurement (green) are the
    same as the BBN-consistent results (red and blue, respectively) in \cref{fig:dNfs_dNfld}.
    The qualitative findings of \cref{fig:relic-tensions} for fluidlike relics extend to
    free-streaming relics when excluding ACT data, but even without LBT, including ACT data
    severely restricts new light relics of either form.
    Results are presented as in \cref{fig:dNfs_dNfld}.
}
\label{fig:relic-tensions-violin}
\end{figure}

As discussed in \cref{sec:dNfs_dNfld}, \Planck{} data allow a greater contribution to $\Neff$ from
self-interacting species, providing a more promising solution to the BAO--CMB and Hubble tensions
than standard free-streaming relics~\cite{Allali:2024cji, Saravanan:2025cyi}.
We reassess fluidlike-relic solutions in \cref{fig:relic-tensions}; to illustrate the model's effect
on tensions, we include SPT-3G data but exclude ACT DR6 CMB data due to the latter's strong
preference for $\Neff < 3.044$ (see \cref{fig:relic-tensions-violin} for comparison and
\cref{sec:new_physics} for discussion).
CMB and DESI BAO data prefer a light relic's reduction in $\Omega_m$ at $\approx 2 \sigma$, though
at the expense of some goodness of CMB fit (as evidenced by the separation of posterior modes fit to
the CMB alone and with DESI).
The same fit prefers a Hubble constant that, though larger than in \LCDM{}, remains in $\approx
3.5 \sigma$ tension with local measurements~\cite{H0DN:2025lyy}.
The LBT measurement, however, severely restricts such larger values of $\Delta \Nfld$, as reported
in \cref{fig:dNfs_dNfld}, preventing fluidlike radiation from alleviating either tension.

\Cref{fig:relic-tensions-violin} shows that the qualitative results of \cref{fig:relic-tensions}
extend to free-streaming radiation; as found in \cref{fig:dNfs_dNfld}, \Planck{}'s greater
compatibility with fluidlike radiation~\cite{Allali:2024cji, Saravanan:2025cyi} is reduced
by including SPT-3G.
The impact of DESI and LBT is similar in each case, since they are mostly (or entirely, for LBT)
sensitive to background effects.
\Cref{fig:relic-tensions-violin} also shows the uniform penalty from including ACT DR6 CMB data,
which independently precludes light relic solutions to cosmological tensions.

\subsubsection{Varying fundamental constants}\label{sec:varying-constants}

\begin{figure}[t]
\includegraphics[width=\textwidth]{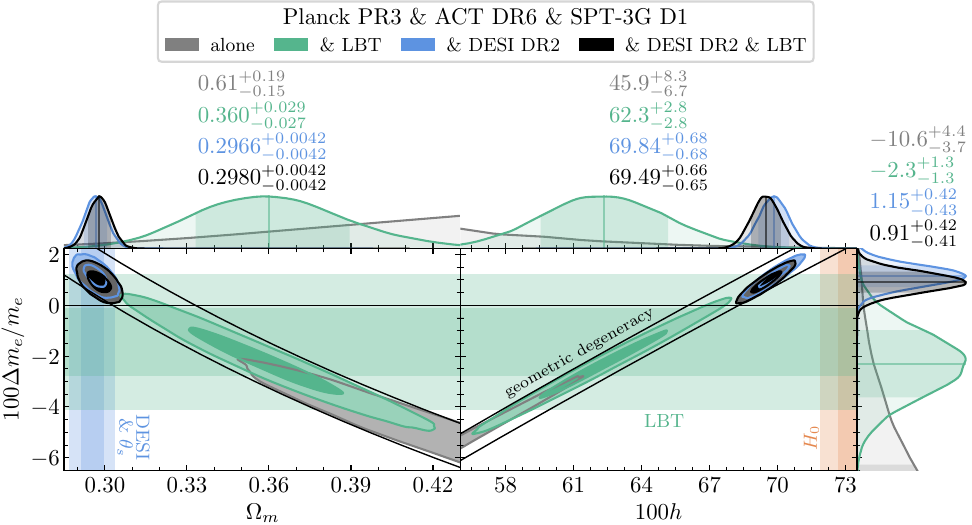}
\caption{
    Impact of LBT's helium abundance measurement on the varying electron mass solution
    to cosmological tensions in minimal scenarios where the variation applies during both
    recombination and nucleosynthesis. 
    An early-time shift $\Delta m_e$ changes the redshift of recombination, opening a geometric
    degeneracy that allows the CMB to be compatible (in principle) with substantially different
    matter fractions $\Omega_m$ or Hubble constants $h$; combining with DESI DR2 data (blue
    contours), the model fully resolves the geometric tension with the acoustic scale data
    (indicated in terms of $\Omega_m$ by the vertical blue band) but not (simultaneously) the
    tension with local determinations of the Hubble constant (vertical orange
    band)~\cite{H0DN:2025lyy}.
    LBT constrains the prerecombination shift (horizontal green band) and independently breaks this
    geometric degeneracy (green contours), yielding parameters more incompatible with low-redshift
    distances than in \LCDM{}.
    However, the helium abundance's low sensitivity to the electron mass yields a comparatively weak
    constraint that has only a small impact on the result from the CMB and DESI (black).
}
\label{fig:me-tensions}
\end{figure}

\begin{figure}[t]
\includegraphics[width=\textwidth]{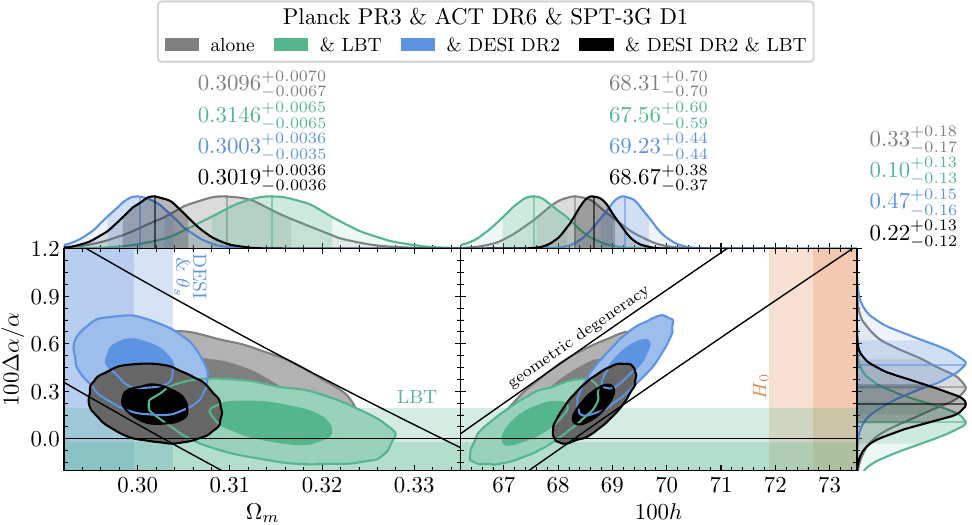}
\caption{
    Same as \cref{fig:me-tensions}, but for a shift $\Delta \alpha$ in the early-time value of the
    fine-structure constant.
    The CMB independently infers a positive shift, a preference whose origin at least partly
    coincides with that of ACT's preference for $\Neff$ below the SM
    value~\cite{AtacamaCosmologyTelescope:2025nti}.
    The CMB fit significantly reduces the tension with DESI BAO data in \LCDM{}, as a larger
    fine-structure constant also realizes early recombination; combining CMB and DESI data yields a
    joint $\approx 2.9 \sigma$ preference.
    As in \cref{fig:me-tensions}, these fits do not come close to resolving the Hubble tension.
    In contrast to the electron mass case, the LBT measurement significantly opposes the preference
    for $\Delta \alpha > 0$, owing to the helium abundance's much greater sensitivity to the 
    fine-structure constant than the electron mass.
}
\label{fig:alpha-tensions}
\end{figure}

We next study the interplay of LBT's helium abundance measurement, constraints on the low-redshift
expansion history from the acoustic scale, and the CMB's geometric degeneracy in varying fundamental
constants scenarios in \cref{fig:me-tensions,fig:alpha-tensions}.
We follow the implementation of Refs.~\cite{Baryakhtar:2024rky, Baryakhtar:2025uxs, Weiner:2026sfm}.
Combining the geometric information in $\theta_s$ with the shape information in CMB anisotropies,
the latter of which constrains relative energy densities at recombination, only fixes the late-time
expansion history in \LCDM{} when the scale factor of recombination $a_\star = T_0 / T_\star$ is
known.
Fixing $\theta_s$ and density ratios at recombination requires the Hubble constant and matter
fraction to scale approximately with $h \propto a_\star^{-3}$ and 
$\Omega_m \propto a_\star^{5}$~\cite{Sekiguchi:2020teg, Baryakhtar:2024rky}, respectively.
Early recombination therefore allows for higher Hubble constants~\cite{Hart:2019dxi,
Sekiguchi:2020teg} and lower matter fractions~\cite{Baryakhtar:2024rky, Weiner:2026sfm} (or the
opposite for late recombination).
A straightforward mechanism to modify $a_\star$ is a shift in fundamental constants: since
$1/a_\star = T_\star / T_0 \propto \alpha_i^{2} m_{e, i}$, with $\alpha_i$ and $m_{e, i}$ the
early-time values of the fine-structure constant and electron mass, the Hubble constant and matter
fraction scale approximately with $h \propto \alpha_i^6 m_{e, i}^3$ and
$\Omega_m \propto \alpha_i^{-10} m_{e, i}^{-5}$.

The geometric degeneracy at play in early recombination is achieved especially well by varying the
electron mass, as it (unlike the fine-structure constant) has no secondary direct effects on the
CMB.
That said, \cref{fig:me-tensions} displays a mild but notable preference for lower early-time
electron masses (and therefore late rather than early recombination) from the CMB alone, indirectly
owing to a preference for large matter fractions that, when not pinned down by $\theta_s$ as in
\LCDM{}, can resolve the CMB lensing excess and low-$\ell$ deficit~\cite{Baryakhtar:2024rky}.
As evident in \cref{fig:alpha-tensions}, recent data from
ACT~\cite{AtacamaCosmologyTelescope:2025blo} separately drive a preference for larger early-time
fine-structure constants~\cite{AtacamaCosmologyTelescope:2025nti}, in particular due to its effects
on small-scale damping~\cite{Kaplinghat:1998ry, Baryakhtar:2024rky}.

As illustrated by \cref{fig:me-tensions,fig:alpha-tensions}, early recombination can essentially
eliminate the tension between the CMB and DESI BAO data~\cite{Baryakhtar:2024rky, Weiner:2026sfm}.
This tension is fully quantified in \LCDM{} by the low matter fraction $\Omega_m$ preferred by
acoustic scale data (combining DESI BAO data with $\theta_s$ from the CMB), but in general it
represents a matter-era distance excess that is mediated by early recombination via the greater
dilution of the matter density between recombination and the redshifts at which DESI measures
distances~\cite{Weiner:2026sfm}.
While the broader geometric degeneracy opened by varying the electron mass would in principle allow
for Hubble constants consistent with local determinations~\cite{H0DN:2025lyy}, the required $m_{e, i}$ far exceeds that allowed by the CMB and DESI.

When early recombination is mechanized by misaligned, hyperlight scalars coupled to electrons
or photons~\cite{Baryakhtar:2024rky, Baryakhtar:2025uxs}, the same shifts in fundamental constants
generally also apply at nucleosynthesis.
The LBT measurement of $\Yp$ opposes the preferences for a larger early-time electron mass (when including DESI) or
fine-structure constant, in particular precluding the varying fine-structure constant's resolution
of the BAO--CMB tension and explanation of ACT's measurements of the damping tail.
The helium fraction increases with both $\alpha_i$ and $m_{e, i}$ approximately as
$\Yp \propto \alpha_i^{2.5} m_{e, i}^{0.4}$~\cite{Baryakhtar:2025uxs}; because the SM
prediction is $1 \sigma$ above the measured value, LBT by itself prefers each to be smaller at early times
by $1 \sigma$.
Because the helium abundance's sensitivity to the electron mass is relatively low ($\sim 0.4$), the
LBT measurement $100 \Delta m_e / m_e = -1.4 \pm 1.3$ (where $\Delta m_e = m_{e, i} - m_e$) has a
rather small effect when added to the CMB and DESI combination.
The fine-structure constant is affected much more significantly (given the sensitivity coefficient
$\sim 2.5$), with the LBT constraint $100 \Delta \alpha / \alpha = -0.23 \pm 0.21$ being competitive
in precision with the CMB's $0.33_{-0.17}^{+0.18}$ and reducing the $2.9\sigma$ preference from the
CMB and DESI to $1.8\sigma$.
These constraints may be translated to the fundamental couplings of hyperlight scalar fields by
following the analysis of Ref.~\cite{Baryakhtar:2025uxs}.

While \cref{fig:me-tensions,fig:alpha-tensions}, like the rest of this work, fixed the neutrino
mass sum to $60~\mathrm{meV}$ (via a single massive species) for simplicity, massive neutrinos have
a nontrivial effect on solutions to the BAO--CMB tension when varied.
In particular, a distinct geometric degeneracy is realized by early recombination combined with a
postrecombination increase in the matter abundance~\cite{Baryakhtar:2024rky, Loverde:2024nfi,
Weiner:2026sfm}, whether from massive neutrinos that become nonrelativistic or the hyperlight scalar
fields whose dynamics explain the variation in constants (but whose gravitational effects we also
neglect in \cref{fig:me-tensions,fig:alpha-tensions} for simplicity; see
Ref.~\cite{Baryakhtar:2024rky}).
Because larger neutrino masses exacerbate the matter-era distance excess~\cite{Weiner:2026sfm}, the
CMB and DESI combination prefers greater shifts in $\alpha_i$ and (especially) $m_{e, i}$ when
marginalized over neutrino masses~\cite{Baryakhtar:2024rky}, limited by the CMB lensing excess's
independent restriction on massive neutrinos' suppression of structure.

\subsubsection{Role of deuterium measurements}\label{sec:deuterium}

\Cref{sec:tensions-radiation,sec:varying-constants} consider the role of measurements of the
helium fraction but not deuterium abundance relative to hydrogen, $\mathrm{D}/\mathrm{H}$.
LBT's helium measurement generally contains the majority of the information from light element
abundances on the types of new-physics scenarios we consider: it is twice as precise as current
$\mathrm{D}/\mathrm{H}$ measurements, and uncertainties in nuclear reaction rates contribute an even
larger theoretical uncertainty in $\mathrm{D}/\mathrm{H}$ predictions~\cite{Fields:2019pfx,
Burns:2026wlw}.
Moreover, differences in modeling of the energy dependence of nuclear reactions currently source
significant discrepancies in deuterium predictions~\cite{Pitrou:2020etk, Yeh:2020mgl,
Pisanti:2020efz, Launders:2026ciu}.
Beyond these cosmology-independent issues, $\mathrm{D}/\mathrm{H}$ is sensitive to the
baryon-to-photon ratio $\eta$ as well as new physics; though in \LCDM{} it therefore provides a
high-redshift anchor to complement the CMB, $\Yp$'s insensitivity to $\eta$ leaves its full
measurement sensitivity to constrain new physics.

With new light degrees of freedom or time-varying fundamental constants, however, the inference of
the baryon abundance from deuterium is altered by its additional dependence on new physics.
For instance, $\mathrm{D}/\mathrm{H} \propto \eta^{-1.6} \cdot \omega_r / \omega_\gamma$ [derived in
the same manner as \cref{eqn:abundance-scalings}].
Likewise, the deuterium prediction increases~\cite{Dent:2007zu,
Burns:2026wlw} with the early-time values $m_{e, i}$ or $\alpha_i$ approximately
as $\alpha_i^{1.4}$ and $m_{e, i}^{0.5}$~\cite{Burns:2026wlw}.\footnote{
    Note that sensitivity coefficients for $\alpha_i$ must combine the intrinsic dependence
    reported in Ref.~\cite{Burns:2026wlw} with the indirect dependence on the neutron-proton mass 
    splitting weighted by the contribution from QED effects~\cite{Thomas:2014dxa, 
    BMW:2014pzb, Coc:2006sx, Bouley:2022eer, Baryakhtar:2025uxs}.
}
These degeneracies in the deuterium prediction require an increase in $\eta$ in response to new
physics that raises the CMB-inferred Hubble constant and are directionally aligned with the CMB's
own intrinsic degeneracies.
A higher baryon density can be required to compensate for modifications to diffusion damping or for
the reduction in the baryon-to-CDM ratio from the increase in CDM density that preserves the onset
of matter domination~\cite{Saravanan:2025cyi}.
Likewise, since the CMB depends on density ratios at recombination~\cite{Eisenstein:2004an,
Hu:2004kn}, it requires a larger present-day baryon density (and therefore larger $\eta$) in
early-recombination models~\cite{Sekiguchi:2020teg, Baryakhtar:2024rky}.

In sum, deuterium measurements have a subdominant role in constraining new physics in the early
Universe that alleviates the Hubble tension.
In contrast, recent work claimed a generic tension between determinations of the baryon density
$\omega_b$ from BBN (through deuterium) and from the CMB in models that alleviate the Hubble
tension~\cite{Giovanetti:2026aku}.
However, the claim only holds in a limited class of scenarios featuring extra energy density at
recombination that is absent at nucleosynthesis like early dark energy~\cite{Poulin:2018cxd}, such that the Standard Model
predictions for $\Yp$ and $\mathrm{D}/\mathrm{H}$ are unmodified.
New light relics are typically populated before nucleosynthesis; specific
scenarios~\cite{Aloni:2023tff, Garny:2024ums} do feature dark radiation that emerges after nucleosynthesis, as was
assumed in Ref.~\cite{Giovanetti:2026aku}, but CMB fits to such models do not independently require
such timing.
Likewise, in high-redshift recombination models (which Ref.~\cite{Giovanetti:2026aku} did not
consider), a deviation from the SM values for the constants that is localized in time around
recombination seems relatively contrived a priori, as it would require transient scalar field
dynamics before and after (but not during) recombination.
Varying constants and extra radiation models therefore generally require a larger baryon density to
preserve the deuterium prediction, like for the CMB.

Deuterium measurements thus do not generically realize a tension in the baryon density in
extra-density models.
The helium fraction is most informative in general, as it is more precisely measured, more uniquely
sensitive to new physics, and not subject to uncertainties and modeling differences in nuclear
reaction rates.
The discrepancy in deuterium predictions from distinct treatments of the energy dependence of
nuclear reactions~\cite{Pitrou:2020etk, Yeh:2020mgl, Pisanti:2020efz, Launders:2026ciu} currently
leads to inconsistent implications for Hubble tension solutions~\cite{Schoneberg:2026vaf}.
For the treatment employed in Ref.~\cite{Giovanetti:2026aku}, the baryon density from deuterium is
in tension with that from the CMB even in \LCDM{}; deuterium measurements contribute almost no
constraining power (in terms of relative information) but rather shift central values as a
compromise in the fit to $\mathrm{D}/\mathrm{H}$ with the fit to the CMB.
More recent deuterium measurements also reduce the tension by $\sim 1
\sigma$~\cite{Kislitsyn:2024jvk, ParticleDataGroup:2026xqw, Poulin:2026ltf} relative to the
measurement used in Ref.~\cite{Giovanetti:2026aku}.

Reference~\cite{Schoneberg:2026vaf} derived analogous results that use LBT's measurement of $\Yp$
(including for the time-varying electron mass scenario), though their presentation focused on
comparisons to the measurement of $\omega_b$ from deuterium in standard cosmology, whereas the
inference of $\omega_b$ is modified in most of the models considered.
The reduction in parameter space due to the addition of $\Yp$ and $\mathrm{D}/\mathrm{H}$
measurements is nearly identical between the two treatments of nuclear reaction rates compared in
Ref.~\cite{Schoneberg:2026vaf}, one of which agrees precisely with CMB measurements of $\omega_b$
when fit to the $\mathrm{D}/\mathrm{H}$ measurement; this similarity indicates that (except for
cases of early dark energy) LBT's helium abundance measurement plays the dominant role.
We confirm the subdominant role of deuterium by importance sampling the joint result from the CMB,
DESI, and LBT presented in \cref{fig:me-tensions}, yielding a shift in $\Delta m_e$ of only a
third of a standard uncertainty (and a $1.5\%$ reduction in uncertainty).
Similar results hold for light relics.

\subsection{Neutron lifetime anomaly}\label{sec:neutron-lifetime}

The neutron lifetime is a key input to the standard model of BBN, since it determines both the rate
of neutron-proton conversion before weak decoupling and the rate of neutron decay into protons.
The free neutron fraction when the deuterium bottleneck clears directly sets the helium fraction,
since effectively all free neutrons are captured into helium nuclei~\cite{Yeh:2023nve}.
The PDG value of the neutron lifetime, $\tau_n = 878.3\pm 0.4~\mathrm{s}$, combines measurements
from multiple ``bottle'' experiments, which measure the lifetime by tracking the number of ultracold
neutrons stored in a trap~\cite{ParticleDataGroup:2026xqw, Musedinovic:2024gms}.
Another approach, of so-called ``beam'' experiments, instead counts the neutron decay products of a neutron beam. 
Beam experiments that measure the proton decay products measure 
$\tau_n = 888.45 \pm 1.65~\mathrm{s}$~\cite{Yue:2013qrc, Rajan:2018hii}, $6 \sigma$ discrepant with
bottle experiments.
If not explained by unknown systematics, this neutron lifetime anomaly may have a BSM
explanation~\cite{Fornal:2018eol,Rajendran:2020tmw,Tan:2019mrj,Cline:2018ami,Berezhiani:2018eds, McKeen:2020oyr,Hostert:2022ntu}, though the SM
prediction for the decay rate is consistent with the bottle measurement~\cite{Czarnecki:2018okw,
Berezhiani:2018udo}.
On the other hand, recent beam experiments that instead observe the electron decay products measure
$\tau_n = 877.2^{+4.4}_{-4.0}~\mathrm{s}$~\cite{Fuwa:2024cdf}, consistent with bottle experiments
and in $\approx 2.6 \sigma$ tension with proton-detection beam experiments.

\begin{figure}[t]
\includegraphics[width=4.5in]{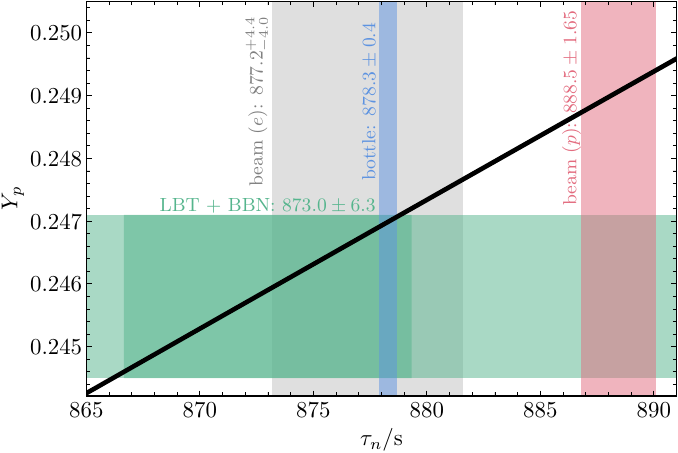}
\caption{
    Measurements of the neutron lifetime $\tau_n$ from a variety of terrestrial experiments,
    compared with the value inferred from LBT's measurement of the helium yield (green) under the
    standard model of BBN (black line).
    Terrestrial experiments include the PDG average value derived from eight bottle experiments
    (blue)~\cite{ParticleDataGroup:2026xqw}, proton-detection based beam experiments
    (red)~\cite{Yue:2013qrc, Rajan:2018hii}, and electron-detection based beam experiments
    (grey)~\cite{Fuwa:2024cdf}.
    The black line depicts the BBN prediction for $\Yp$ as a function of $\tau_n$, with width
    corresponding to the $\pm 1 \sigma$ interval in $\eta$ as inferred from \Planck{}, SPT, and ACT
    CMB data in \LCDM{}.
    The BBN-based result is consistent to $1 \sigma$ with measurements from the bottle and
    electron-detection beam experiments and is $2.4 \sigma$ lower than that from proton-detection
    beam experiments.
}
\label{fig:neutron_lifetime}
\end{figure}

Primordial helium abundance measurements provide an independent probe of the neutron
lifetime~\cite{Chowdhury:2022ahn, Yeh:2023nve}.
Instead of taking lifetime measurements from bottle experiments, the default choice in BBN analyses,
we may invert measurements of the helium abundance to constrain $\tau_n$~\cite{Chowdhury:2022ahn,
Yeh:2023nve}.
Given the extremely weak dependence of $\Yp$ on the baryon-to-photon ratio, this extraction is
robust to whatever independent information (CMB or deuterium observations) is used to constrain
$\eta$; the only assumption required is that no new physics modifies BBN. 
\Cref{fig:neutron_lifetime} shows\footnote{
    We simply model the $\tau_n$ dependence as $\propto \tau_n^{0.73}$~\cite{Yeh:2023nve}, which is 
    entirely sufficient to keep errors below a tenth of LBT's uncertainty over the 
    $\approx 3\%$ range shown in \cref{fig:neutron_lifetime}.
} that LBT measures $\tau_n = 873.0 \pm 6.3~\mathrm{s}$, commensurate
with bottle experiments and the electron-detection beam experiment and discrepant with
proton-detection beam experiments' measurement at $2.4 \sigma$.

\section{Conclusions}\label{sec:conc}

In this paper, we studied the impact of the Large Binocular Telescope project's recent measurement
of the primordial helium abundance, when combined with the CMB, on constraining cosmological and
inflationary parameters, searching for new physics in the radiation sector, and addressing
cosmological tensions and the neutron lifetime anomaly.
The helium abundance's importance to CMB anisotropies stems from its determination of the free
electron fraction, through which it directly affects recombination and the
diffusion damping of anisotropies on small scales.
The most precise cosmological inference from the CMB therefore assumes the cosmological parameter
dependence of $\YHe$ according to the standard theory of BBN.
Relaxing this assumption degrades the CMB's measurements of the baryon density, the abundance of light
relics, and the scalar spectral index by order-unity factors.

\Cref{sec:cosmo} demonstrates that empirically calibrating $\YHe$ with the LBT measurement
recovers the constraining power otherwise lost by being agnostic to the physics of BBN.
\Cref{fig:diff_models_PAS} shows that the empirical calibration recovers measurements of $\omega_b$
and $\Neff$ that are identical to those that instead enforce BBN consistency, and
\cref{fig:r_n_s_PAS} shows that this conclusion extends to inflationary parameters that determine
the shape of the primordial power spectrum.
In \cref{sec:forecasts}, we forecast for future CMB surveys that the LBT measurement is already
nearly sufficient for maximally precise BBN-agnostic constraints on $\Neff$ as well as the spectral
tilt and its running; a factor of $\sim 2$ improvement in the empirical measurement would suffice
for BBN-agnostic constraints on \LCDM{} parameters to match BBN-consistent ones.

An empirical calibration of $\YHe$ can fully substitute for a theoretical one when
constraining $\omega_b$ and inflationary parameters because the predicted helium abundance is only
very weakly sensitive to the baryon-to-photon ratio [per \cref{eqn:abundance-scalings}] and is
entirely insensitive to inflationary parameters.
On the other hand, the theoretical prediction of the helium abundance also depends directly on
$\Neff$, making empirical and theoretical calibrations complementary; imposing both yields the
tightest available determination.
The central value of $\Neff$ nonetheless depends on which combination of CMB and BAO datasets is
employed, as examined in \cref{sec:new_physics,sec:full_constraints} and discussed below.

In \cref{sec:new_physics}, we exploit the joint power of modeling BBN and CMB simultaneously to constrain nonminimal extensions to the radiation sector. 
We first studied additional radiation beyond the SM neutrinos that is either
free-streaming (as neutrinos are) or fluidlike due to efficient self-interactions.
\Cref{fig:dNfs_dNfld} shows that including ACT DR6 in addition to \Planck{} PR3 and SPT-3G D1 CMB
data drives more stringent upper limits on fluidlike than on free-streaming radiation.
ACT therefore distinguishes between the effects of their perturbations, since equal amounts of either
type of light relic affect the expansion history identically (see
\cref{fig:dNfs_dNfld_A_s_theta_s_lite} in \cref{sec:full_constraints} for a more detailed analysis).
When employing both theoretical and empirical calibration, however, the upper limits on additional
free-streaming and fluidlike radiation are nearly identical, demonstrating that
background-level information dominates over the distinguishing signatures from perturbations.
ACT thus tightens these limits mostly because it prefers less damping than predicted in \LCDM{} and therefore a radiation density nearly $2 \sigma$ below the SM prediction.

We then performed a phenomenological test that more generally established the primacy of background-level effects by deriving constraints that allow an arbitrary mixture of free-streaming and fluidlike radiation.
This parametrization captures a wide range of radiation-sector behavior while enabling a more
``apples-to-apples'' comparison of CMB constraints with those from BBN alone, since primordial
element abundances are insensitive to perturbations in the radiation sector.
\Cref{fig:violin_plot_N_x_PS_PAS_lens} shows that ACT again drives the radiation density below the SM prediction,
regardless of the input for $\YHe$ or of marginalization over the composition of the radiation sector.

The combination of precise CMB and light element abundance measurements enables searches for
evolution in the radiation sector between nucleosynthesis ($\NeffBBN$) and photon decoupling
($\NeffCMB$), as studied in \cref{sec:bbn_vs_cmb}.
Although \Planck{} and SPT data are concordant with Standard Model
expectations across a wide range of modifications to the radiation sector, ACT's preference for less damping drives $\NeffCMB$ below its
Standard Model value, even as the LBT measurement finds no such departure at nucleosynthesis.
Microphysical models that realize this post-BBN reduction in the radiation density require nontrivial
modifications to SM physics, including effects beyond the phenomenological test we
performed~\cite{Escudero:2026mgw}.

The outsized impact of ACT's damping tail is a common theme of our searches for new light relics,
but DESI's baryon acoustic oscillation data partially counteract the downward shift in $\Neff$.
In fact, \cref{sec:tensions-radiation} showed that combining DESI with \Planck{} and SPT CMB data
yields an $\approx 2 \sigma$ preference for new light relics, as they can indirectly alleviate the
BAO--CMB tension through changes to density ratios inferred at last
scattering~\cite{Saravanan:2025cyi}.
However, LBT's consistency with the SM excludes explanations through typical models of light relics
(see \cref{fig:relic-tensions-violin}), requiring extensions to evade their effect on
BBN~\cite{Aloni:2023tff}; either way, ACT precludes a light-relic solution outright.
As a corollary, the improved agreement with the SM prediction when all CMB data are combined with
DESI (as in \cref{fig:diff_models_PAS}) merely reflects a compromise between the two competing dataset
preferences---one that would not persist if the BAO--CMB tension were otherwise addressed~\cite{Weiner:2026sfm}.
Likewise, the greater consistency of $\Neff$ with the SM when combining CMB and LBT data masks the improvement in fit to ACT data from reducing the degree of damping.

In \cref{sec:cosmo-tensions}, we further assess the impact of the LBT measurement on proposed explanations of the Hubble tension and the BAO--CMB tension.
Precision measurements of the helium fraction are uniquely poised to independently test the wide variety of proposals that also predict effects on BBN, including not just light relics but also time-varying fundamental constants.
The LBT measurement opposes the time-variation in the fine-structure constant
(\cref{fig:alpha-tensions}) preferred by CMB data both independently and combined with
DESI~\cite{AtacamaCosmologyTelescope:2025nti, Weiner:2026sfm}.
The helium fraction is much less sensitive to variations in the electron mass, on the other hand,
leaving it a viable early-recombination mechanism to address the BAO--CMB tension~\cite{Baryakhtar:2024rky,Baryakhtar:2025uxs,Weiner:2026sfm}.
None of these models resolves the Hubble tension.
We also commented on the subdominant role of deuterium abundance measurements in the typical
scenario where new physics at recombination also affects BBN (as characteristic to the models we
considered) in \cref{sec:deuterium}.
Finally, in \cref{sec:neutron-lifetime} we invert the LBT measurement to infer the neutron lifetime
via its strong impact on the helium fraction, finding results consistent with bottle and electron-detection beam
experiments and $2.4 \sigma$ below proton-detection beam measurements.

Big bang nucleosynthesis anchors the success of the Standard Model in cosmology at high temperature,
with its prediction for the helium fraction both key to modeling the cosmic microwave background at
lower temperature and now independently confirmed below the percent level through astrophysical
observations~\cite{Aver:2026dxv}.
LBT's empirical precision enables robustness tests of the cosmological model as fit to both
present and future CMB data and can play a decisive role in testing candidate explanations of
cosmological tensions.
Moreover, it provides an agnostic route to measuring cosmological parameters with near-optimal
precision without modeling nucleosynthesis at all.
One intriguing application is current CMB data's preference for a radiation density below the SM
prediction for neutrinos that, aside from presenting a challenge in modeling, by default assumes an
unphysical extrapolation to a decrement in the expansion rate alone at BBN.
Astrophysical measurements of primordial abundances stand to provide invaluable empirical and
theoretical cross checks on the various extant curiosities of the post-BBN Universe.

\begin{acknowledgments}
We acknowledge helpful discussions with Masha Baryakhtar, Cara Giovanetti, Ella Henry, Junwu Huang, Jessie Shelton, and Nikita A. Zemlevskiy.
\Cref{fig:r_n_s_PAS} was made with the help of the code available at~\cite{balkenhol2025rns}.
MMS and ML are supported by the Department of Energy grant DE-SC0011637 and the Dr. Ann Nelson Endowed Professorship. 
Research at Perimeter Institute is supported in part by the Government of Canada through the Department of Innovation, Science and Economic Development and by the Province of Ontario through the Ministry of Colleges and Universities.
This work was enabled, in part, by the use of advanced computational, storage, and networking infrastructure provided by the Hyak supercomputer system at the University of Washington~\cite{uwhyak}, which was supported by the UW Student Technology Fee, the UW Department of Physics, and the College of Arts and Sciences. 
We acknowledge the use of various \texttt{Claude} large language models by Anthropic for orchestrating batch jobs for MCMC runs, refining plotting code, assisting in literature search, and proofreading.
This work made use of a number of open-source software packages~\cite{Foreman-Mackey:2012any,
Hogg:2017akh, Foreman-Mackey:2019, Harris:2020xlr, Virtanen:2019joe, Hunter:2007ouj,
hoyer2017xarray}.
\end{acknowledgments}

\appendix

\section{Supplementary results}\label{sec:full_constraints}

\begin{figure}[t]
    \centering
    \includegraphics[width = \textwidth]{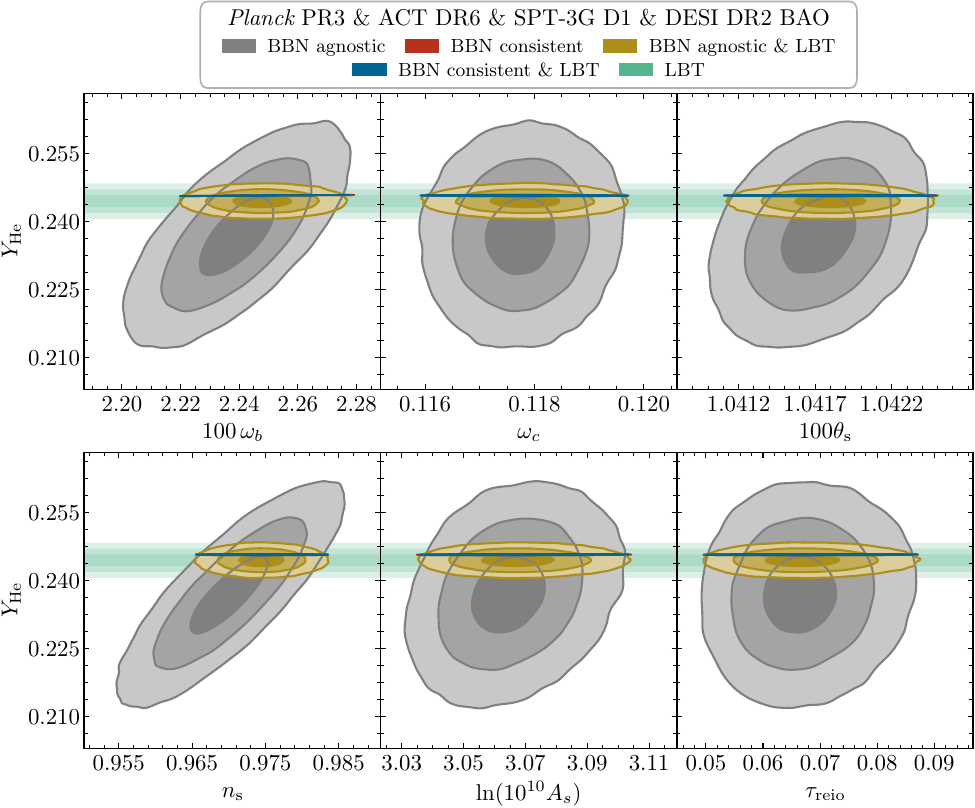}
    \caption{
        Joint posterior distributions of the primordial helium mass fraction $\YHe$ with $\omega_b$, $\omega_c$, $\theta_s$, $n_s$, $\ln(10^{10}A_s)$, and $\taureio$ in the flat \LCDM{} model. 
        Results are presented as in \cref{fig:diff_models_PAS}.
    }
    \label{fig:lcdm_PAS}
\end{figure}

In this appendix, we present various extended versions or variants of figures from
\cref{sec:cosmo,sec:new_physics}.
\Cref{fig:lcdm_PAS} presents the joint posterior of $\YHe$ with all \LCDM{} parameters (as in the left panel of \cref{fig:diff_models_PAS}).
Within \LCDM{}, $\YHe$ is degenerate with the spectral tilt $n_s$ of the primordial power spectrum because $\YHe$ sets the diffusion damping rate.
Other than $\omega_b$ (as also shown in \cref{fig:diff_models_PAS}) and $n_s$, only $\theta_s$ exhibits a (weak) correlation with $\YHe$.

\begin{figure}[t]
    \centering
    \includegraphics[width = \textwidth]{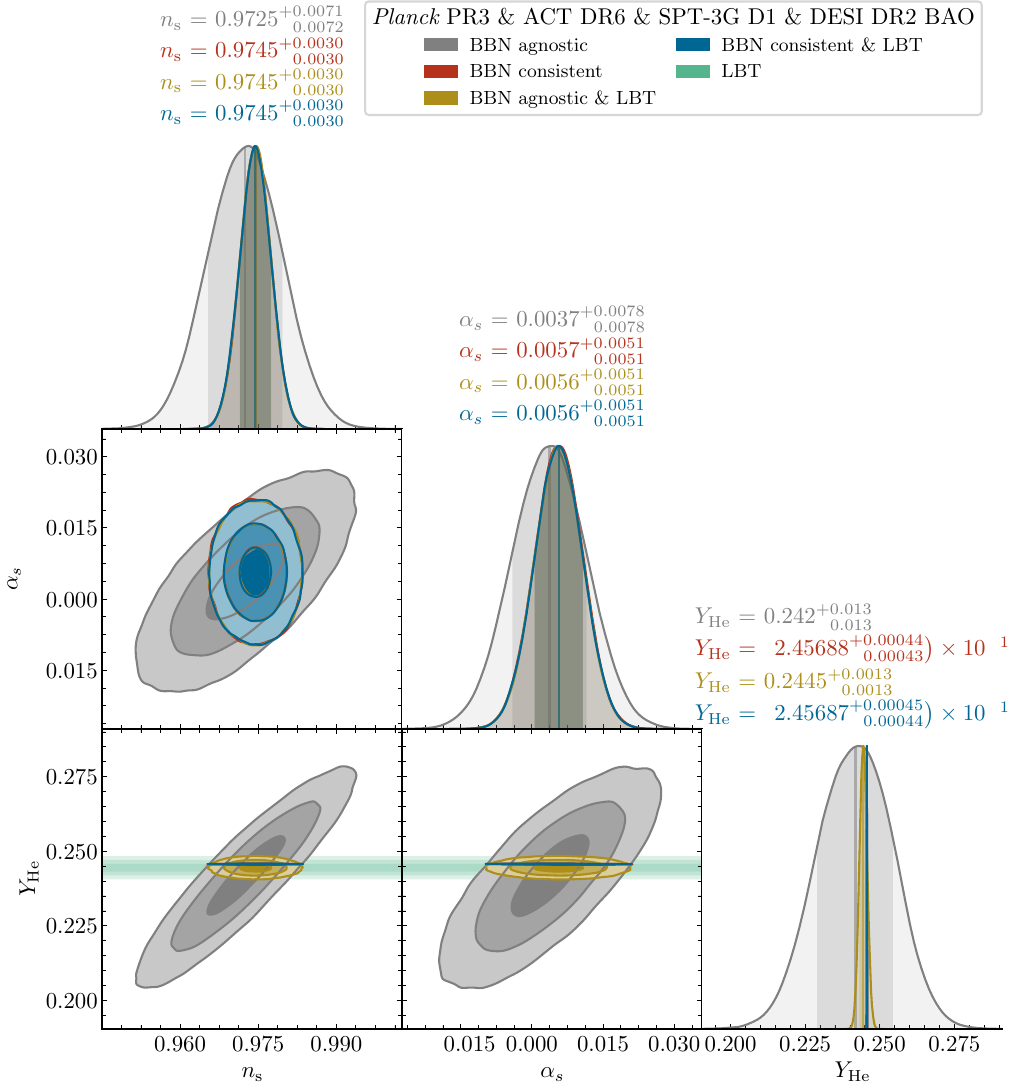}
    \caption{
        Joint posterior distributions of $\YHe$ with the spectral tilt $n_s$ of the primordial power spectrum and its running $\alpha_s$.
        Different colors represent the different modeling choices for $\YHe$ as labeled in the legend. Results are presented as in \cref{fig:diff_models_PAS}. 
        The strong three-way degeneracy between $n_s$, $\alpha_s$, and $\YHe$ in the BBN-agnostic scenario disappears when $\YHe$ is empirically or theoretically calibrated.}
    \label{fig:alpha_PAS}
\end{figure}

\begin{figure}[t]
    \centering
    \includegraphics[width = \textwidth]{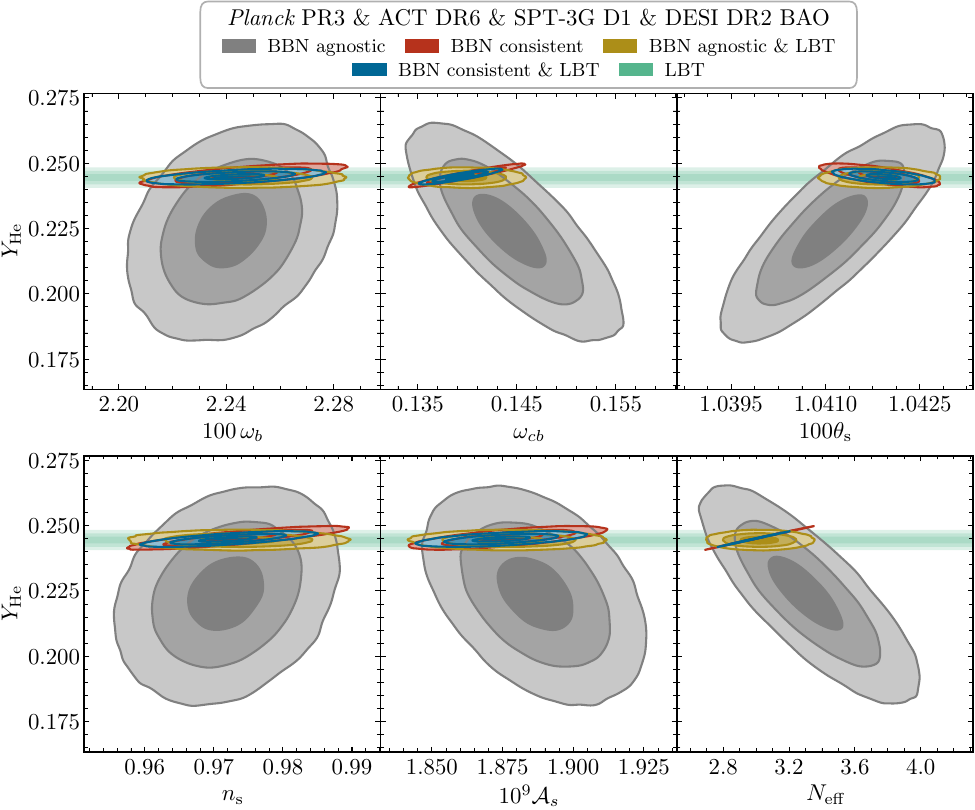}
    \caption{
        Joint posterior distributions of $\YHe$ with $\omega_b$, $\omega_{\mathrm{cb}} \equiv \omega_c + \omega_b$, $\theta_s$, $n_s$, $\mathcal{A}_s \equiv \Astau$, and $\Neff$ in models where the radiation density of free-streaming neutrinos is allowed to vary.
        Results are presented as in \cref{fig:diff_models_PAS}. 
        LBT's measurement of $\YHe$ is larger than the CMB's by more than $1\sigma$. 
        This offset results in minor but not insignificant shifts in the central values of parameters that are degenerate with $\YHe$ (namely $\Neff$, $\omega_{\mathrm{cb}}$, and $\theta_s$) between BBN-agnostic analyses and those that enforce BBN consistency and/or use the LBT measurement.}
    \label{fig:Neff_PAS}
\end{figure}

Just as with $n_s$ in \LCDM{}, a running $\alpha_s$ of the spectral index can compensate for changes in the damping tail, as exhibited by the steep degeneracy between the three parameters in BBN-agnostic analyses in \cref{fig:alpha_PAS}.
Empirically calibrating $\YHe$ with the LBT measurement breaks these degeneracies to the same extent as using standard BBN predictions; combining theoretical and empirical calibration does not improve the precision of any of the parameters in \LCDM{} because $\YHe$ varies quite slowly with $\omega_b$ [\cref{eqn:abundance-scalings}].
On the other hand, models that vary $\Neff$ do benefit from combining theoretical and empirical calibration, as demonstrated in \cref{fig:Neff_PAS}.
ACT's general preference for less damping than \LCDM{} predicts shifts the central values of not
only $\Neff$ but also other parameters when empirical and/or theoretical calibration is enforced.

\begin{figure}[t]
    \centering
    \includegraphics[width = \textwidth]{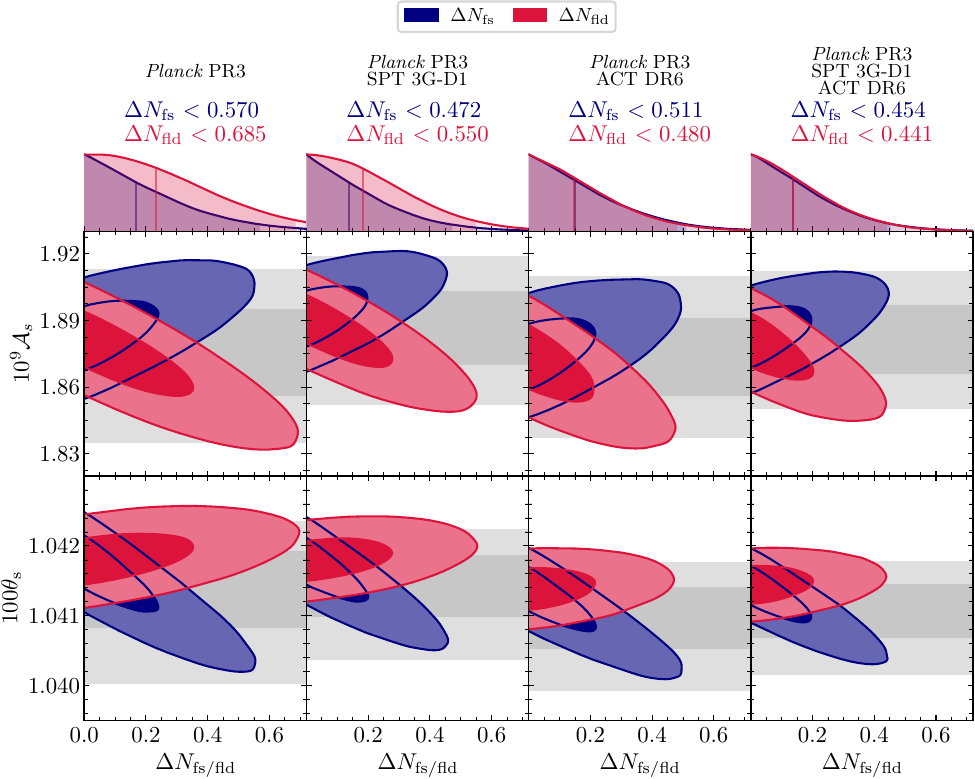}
    \caption{
        Joint posterior distributions of $\theta_s$ and $\mathcal{A}_s \equiv \Astau$ with additional free-streaming ($\Delta \Nfs$, blue) or fluidlike ($\Delta \Nfld$, red) radiation for CMB dataset combinations by column as labeled. 
        Results that include ACT data also cut high-$\ell$ \Planck{} data (see \cref{sec:methods}).
        No lensing data are used and $\YHe$ is varied as a free parameter.
        Grey bands indicate the $1$ and $2 \sigma$ limits on $\theta_s$ and $\mathcal{A}_s$ when allowing for an arbitrary mixture of extra fluidlike and free-streaming radiation (to marginalize over the effects of their perturbations).
        Including ACT data constrains additional fluidlike radiation more strongly than free-streaming radiation due to the upper limit ACT places on $\theta_s$.}
    \label{fig:dNfs_dNfld_A_s_theta_s_lite}
\end{figure}

\Cref{fig:dNfs_dNfld_A_s_theta_s_lite} isolates the effect of each CMB dataset on BBN-agnostic
constraints on additional free-streaming and fluidlike radiation.
Although fluidlike radiation's phase-shift signature is partly canceled by the correlated change in
the fraction of pressure-supported matter~\cite{Saravanan:2025cyi} (as evident in the weaker degeneracy between fluidlike radiation and $\theta_s$), the residual effect is still absorbed
by $\theta_s$; the tighter upper limit ACT places on $\theta_s$ therefore restricts additional
fluidlike radiation more severely than do \Planck{} and SPT data.
SPT, in turn, disfavors lower values of $\Astau$, which penalizes fluidlike relics.

\begin{figure}[t]
    \centering
    \includegraphics[width = \textwidth]{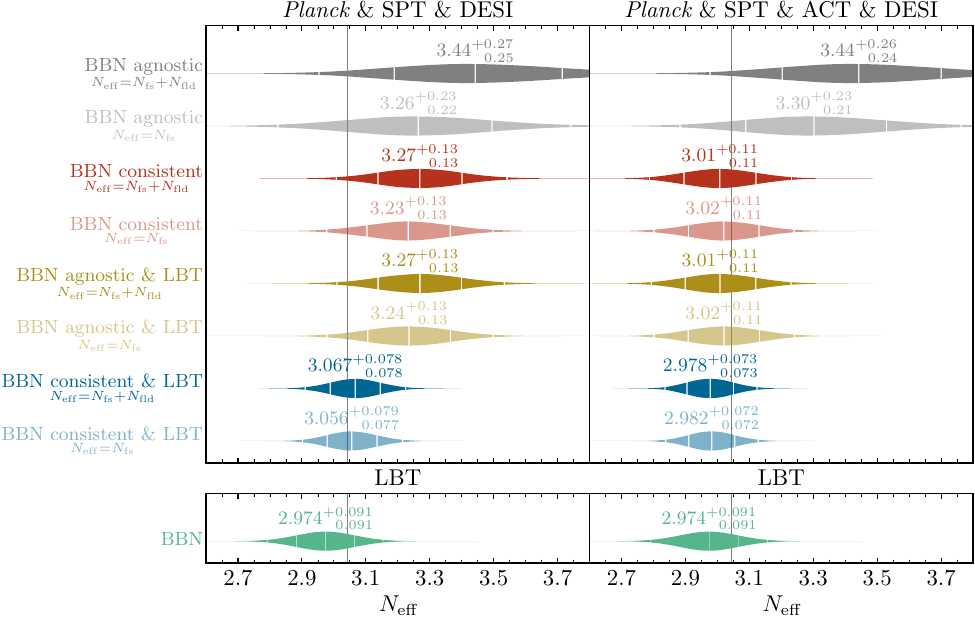}
    \caption{ 
        Measurements of $\Neff$ as in \cref{fig:violin_plot_N_x_PS_PAS_lens}, but including DESI DR2 BAO data, which drive a
        preference for additional radiation (\cref{sec:tensions-radiation}).}
    \label{fig:violin_plot_N_x_PS_PAS_bao_lens}
\end{figure}

\begin{figure}[t]
    \centering
    \includegraphics[width = \textwidth]{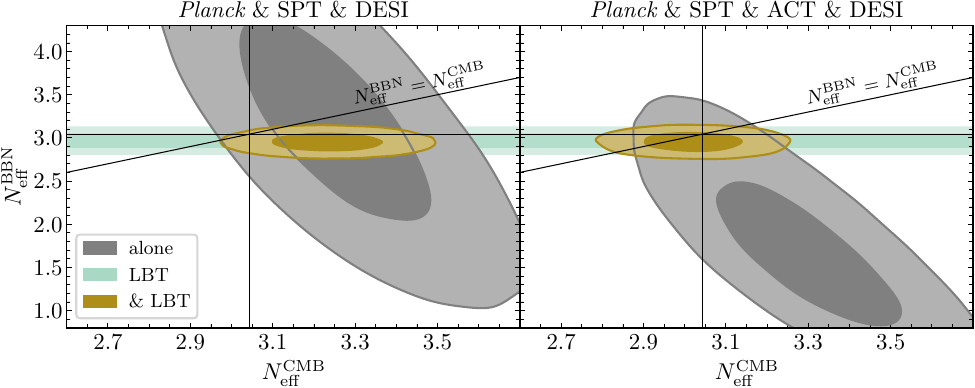}
    \caption{Constraints on the evolution of radiation between nucleosynthesis and recombination as presented in \cref{fig:deltaN}, but including the DESI DR2 BAO data.}
    \label{fig:deltaN_bao}
\end{figure}

DESI's preference for additional radiation counteracts ACT's preference for less radiation regardless of marginalization over the free-streaming fraction, as exemplified in \cref{fig:violin_plot_N_x_PS_PAS_bao_lens}.
Likewise, \cref{fig:deltaN_bao} shows that, even as the LBT measurement pins the value of $\NeffBBN$
to the SM prediction, DESI shifts the inferred value of $\NeffCMB$ upwards; concordance with the SM
depends on which CMB datasets are used.
As emphasized in \cref{sec:standard_cosmo,sec:tensions-radiation}, the concordance of $\NeffCMB$
with the SM prediction when including both ACT and DESI data reflects a compromise between (rather
than a mediation of) the competing effects of ACT's damping tail preferences and the BAO--CMB
tension; independent explanations of either could qualitatively alter these conclusions.

\section{Datasets and methodology}\label{sec:methods}

We include measurements of the CMB temperature and polarization anisotropies from the \Planck{} satellite, the Atacama Cosmology Telescope (ACT), and the South Pole Telescope (SPT) in our analysis.
In particular, we use the \Planck{} 2018 (PR3) high-$\ell$ (TT, TE, and EE) and low-$\ell$ TT data via the nuisance-marginalized likelihoods~\cite{Planck:2018vyg,Planck:2019nip} and the \texttt{Sroll2} low-$\ell$ EE likelihood~\cite{Pagano:2019tci}.
We use the ACT DR6 temperature and polarization foreground-marginalized likelihoods~\cite{AtacamaCosmologyTelescope:2025vnj,AtacamaCosmologyTelescope:2025blo,AtacamaCosmologyTelescope:2025nti,Beringue:2025bur} and the SPT-3G D1 temperature and polarization foreground-marginalized likelihoods~\cite{SPT-3G:2025bzu,Balkenhol:2024sbv}, including the recommended priors on the necessary nuisance parameters.
As in the ACT DR6 analysis~\cite{AtacamaCosmologyTelescope:2025blo,AtacamaCosmologyTelescope:2025nti}, when combining \Planck{} and ACT likelihoods we restrict the \Planck{} likelihoods to $\ell < 1000$ for TT and $\ell < 600$ for TE and EE.
Moreover, we truncate the ACT likelihood at $\ell = 4000$ as there is little cosmological
information at higher multipoles~\cite{Costa:2025kwt}.
We include lensing data from \Planck{}, ACT, and SPT in our analysis as well~\cite{Carron:2022eyg,ACT:2023dou,ACT:2023kun,ACT:2023ubw,SPT-3G:2025bzu,ACT:2025qjh}. At times, we include the latest BAO measurements from the Dark Energy Spectroscopic Instrument (DESI) survey~\cite{DESI:2025zgx,DESI:2025zpo}.
When constraining the tensor-to-scalar ratio, we also include the BICEP/\textit{Keck} B-mode CMB likelihood~\cite{BICEP:2021xfz}.

To carry out our analyses, we use the Boltzmann
code \class{} 3.3.4~\cite{Blas:2011rf,Lesgourgues:2011re,Lesgourgues:2011rh} interfaced with \cobaya{} 3.5.1~\cite{Torrado:2020dgo,2019ascl.soft10019T}.
When including the SPT or ACT datasets, we use a minimal set of precision settings for \class{} as detailed in Appendix D of Ref.~\cite{Costa:2025kwt}.
When enforcing BBN consistency, we use the \texttt{sBBN\_2025\_primat.dat} table included in \class{}, which is computed with the PRIMAT BBN code~\cite{Pitrou:2018cgg,Pitrou:2020etk}.
We model additional free-streaming or fluidlike radiation with the ultrarelativistic fluid species (parametrized by \texttt{N\_{ur}}) implemented in \class{}.
For fluidlike radiation, we set the effective sound speed squared and effective viscosity parameter to $c_{\mathrm{eff, ur}}^2 = 1/3$ and $c_{\mathrm{visc, ur}}^2 = 0$, respectively, which is a standard parametrization for a perfect fluid~\cite{Hu:1998kj}. 
Note that the ultrarelativistic fluid approximation, which is implemented in \class{} to optimize the modeling of free-streaming radiation at late times, assumes that $c_{\mathrm{visc, ur}}^2 = 1/3$. This approximation must therefore be disabled (by setting the parameter \texttt{ur\_fluid\_approximation} to 3) to accurately model a perfect fluid. We model the standard neutrino content with the implementation of noncold dark matter species (\texttt{N\_{ncdm}}).

With the exception of the results in \cref{sec:varying-constants} (which use
\textsf{emcee}~\cite{Foreman-Mackey:2012any,Hogg:2017akh,Foreman-Mackey:2019} and likelihood and
prior implementations as described in Refs.~\cite{Baryakhtar:2024rky,Costa:2025kwt}),
we sample posteriors using the \cobaya{} implementation of the Metropolis-Hastings method for Markov
chain Monte Carlo~\cite{Lewis:2013hha}.
Denoting a uniform distribution between $a$ and $b$ as $\mathcal{U}(a, b)$, we take flat priors for \LCDM{} parameters: $100\theta_s\sim\mathcal{U}(0.5, 10)$, $\omega_b\sim\mathcal{U}(0.005, 0.1)$, $\omega_c\sim\mathcal{U}(0.001, 0.99)$, $\ln \left(10^{10}A_s \right)\sim\mathcal{U}(1.61, 3.91)$, $n_s\sim\mathcal{U}(0.8, 1.2)$, and $\taureio\sim\mathcal{U}(0.01, 0.8)$.
When not otherwise fixed, we take $\YHe \sim \mathcal{U}(0.01, 0.5)$.
The total amount of radiation and the free-streaming fraction are sampled via
$\Neff \sim \mathcal{U}(2, 4.5)$ and the ratio $\Nfld/\Neff \sim \mathcal{U}(0, 1)$.
When searching for additional free-streaming or fluidlike radiation, we sample $\Delta \Nx \sim \mathcal{U}(0, 1)$.
For the analysis in \cref{sec:bbn_vs_cmb}, we sample $\NeffBBN$ with a uniform prior $\mathcal{U}(0.044, 10.044)$ set by the limits of the BBN table.
All posteriors presented contain between $7,000$ and $25,000$ independent samples.

\bibliography{references,manual}

\end{document}